\documentclass[12pt]{iopart}

 \expandafter\let\csname equation*\endcsname\relax
  \expandafter\let\csname endequation*\endcsname\relax
\usepackage{amsmath}
\usepackage{iopams}

\usepackage[T1]{fontenc}
\usepackage[latin9]{inputenc}
\usepackage{geometry}
\usepackage{mathrsfs}
\usepackage{bm}
\usepackage{amsmath}
\usepackage{amssymb}
\usepackage{graphicx}
\usepackage{esint}

\usepackage{epsfig}
\usepackage{graphicx,color}

\newcommand{\beq}{\begin{equation}}
\newcommand{\eeq}{\end{equation}}

\begin{document}

\review[Extreme Plasma Astrophysics]{Plasma Physics of Extreme Astrophysical Environments}
\author{ Dmitri A Uzdensky and Shane Rightley} 
\address{Center for Integrated Plasma Studies, Physics Department, University of Colorado, 
UCB 390, Boulder, CO 80309-0390, USA}
\vspace{-10 pt}
\ead{\mailto{uzdensky@colorado.edu}, \mailto{shane.rightley@colorado.edu}}


\vspace{-10 pt}

\begin{abstract} 
Among the incredibly diverse variety of astrophysical objects, there are some that are characterized by very extreme physical conditions not encountered anywhere else in the Universe.  Of special interest are ultra-magnetized systems that possess magnetic fields exceeding the critical quantum field of about 44 teragauss. There are basically only two classes of such objects: magnetars, whose magnetic activity is manifested, e.g., via their very short but intense gamma-ray flares, and central engines of supernovae and gamma-ray bursts --- the most powerful explosions in the modern Universe. 
Figuring out how these complex systems work necessarily requires understanding various plasma processes, both small-scale kinetic and large-scale magnetohydrodynamic (MHD), that govern their behaviour. 
However, the presence of an ultra-strong magnetic field modifies the underlying basic physics to such a great extent that relying on conventional, classical plasma physics is often not justified. Instead, plasma-physical problems relevant to these extreme astrophysical environments call for constructing relativistic quantum plasma physics based on quantum electrodynamics (QED). 
In this review, after briefly describing the astrophysical systems of interest and identifying some of the key plasma-physical problems important to them, we survey the recent progress in the development of such a theory.  We first discuss the ways in which the presence of a super-critical field modifies the properties of vacuum and matter and then outline the basic theoretical framework for describing both non-relativistic and relativistic quantum plasmas.  
We then turn to some specific astrophysical applications of relativistic QED plasma physics relevant to magnetar magnetospheres and to central engines of core-collapse supernovae and long gamma-ray bursts.  Specifically, we discuss 
the propagation of light through a magnetar magnetosphere; 
large-scale MHD processes driving magnetar activity and responsible for jet launching and propagation in gamma-ray bursts;
energy-transport processes governing the thermodynamics of extreme plasma environments; 
micro-scale kinetic plasma processes important in the interaction of intense electric currents flowing through a magnetar magnetosphere with the neutron star surface; 
and magnetic reconnection of ultra-strong magnetic fields. 
Finally, we point out that future progress in applying relativistic quantum plasma physics to real astrophysical problems will require the development of suitable numerical modeling capabilities.  
\end{abstract}

\vspace{-20 pt}
\pacs{52.25.Xz, 52.27.Ep, 52.27.Ny, 52.30.Cv, 52.35.Vd, 95.30.Qd, 97.60.Jd, 98.70.Rz}

\vspace{-20 pt}
\submitto{\RPP}

\vspace{-10 pt}

\maketitle



\section{Introduction}
\label{sec-intro}

\subsection{The scope and structure of plasma astrophysics}
\label{subsec-scope}

Astrophysical and space plasmas are complex~\cite{Kulsrud-2005}. 
To start with, they are almost always immersed in magnetic fields.  In addition, they are often bathed in radiation fields with photons over a wide range of wavelengths across the electromagnetic spectrum.  Both magnetic and radiation fields can be produced locally by the plasma itself or have an external origin (i.e., represent an external background). 

Next, it is rather rare that one finds a completely quiescent, homogeneous plasma in a complete mechanical and thermal equilibrium.  Quite often, cosmic plasmas have some level of turbulence, i.e., a range of collective motions and structures, including eddies and various kinds of waves,%
\footnote{Here we distinguish between collective coherent plasma waves and incoherent electromagnetic radiation described as individual photons.  A rough criterion for this distinction is whether the photon frequency is higher or lower than the plasma frequency.} 
on a broad range of scales, from the global system size down to dissipative scales (e.g., plasma microscopic scales, such as the ion Larmor radius $\rho_i$ or the ion collisionless skin depth~$d_i$).  Turbulent motions may sometimes play an important dynamical role, i.e., exert pressure and carry energy. 

Furthermore, in sufficiently collisionless plasmas, and especially if small-scale phenomena are involved, the electrons and ions (or positrons) do not have enough time to relax to a complete thermodynamic equilibrium, and hence they often have complex, multi-component distribution functions, e.g., a Maxwellian with a high-energy power-law tail.  For example, in solar wind studies  one often describes the plasma distribution function in terms of three components: the core, the halo, and the strahl~\cite{Pierrard_etal-1999}. 
In some extreme cases, e.g., in Pulsar Wind Nebulae (PWNe), the thermal component is believed to be essentially absent completely and the plasma is characterized by a power-law distribution with a high-energy cutoff, e.g., \cite{deJager_Harding-1992}.  Another important example of high-energy nonthermal particles present in the Interstellar Medium (ISM) are relativistic Cosmic Rays (CRs), characterized by a power-law energy distribution that extends up to~$10^{20}$~eV. 

Finally, the chemical composition of cosmic plasmas can be quite diverse and complex. It can range from pure electron-positron plasmas at highest energies, e.g., in pulsar winds and nebulae, to usual electron-ion plasmas, to cold weakly ionized plasma with a whole host of ion and neutral species, e.g., in cold molecular clouds or in protostellar accretion disks. An important additional complication in the case of cold, partially ionized plasmas is that their proper treatment often requires taking into account a complex network of chemical and photo-chemical reactions. Furthermore, this chemistry and, especially, ionization-recombination balance, is often greatly affected by ubiquitous dust grains of different sizes.  
At the other end of the energy-density spectrum, nuclear chemistry and neutrino processes become important in dense and hot plasmas in central engines of Supernovae (SNe) and Gamma-Ray Bursts (GRBs).

These remarks lead us to conclude that one cannot adequately  describe real space- and astrophysical plasmas with just two or three basic parameters, such as density and temperature, as it is often done in simplified textbook problems.  Instead, one sees that these complex plasmas should be regarded as consisting of several separate but interpenetrating and actively interacting {\it components} or {\it constituents}.  One can list them as follows:  the thermal gas of electrons, ions, and perhaps positrons; a population of nonthermal (e.g., described by a truncated high-energy power-law tail) particles (e.g., cosmic rays); a large-scale (background) magnetic field; electromagnetic radiation (photons); small-scale collective plasma waves or eddies; and finally, in certain cases, neutral atoms and molecules, and dust grains.

Importantly, all these constituents carry energy. The relative importance of the different plasma constituents in the overall energetics can, of course, vary from system to system.  Interestingly, however, one often finds herself in a situation where at least several of the constituents are energetically non-negligible. The ISM in our own Galaxy is a good example (see, e.g., \cite{Ferriere-2001} for a review): here, remarkably, the energy densities (and hence pressures) of the thermal gas, of Cosmic Rays, of the background magnetic field, of light, and of turbulence are all generally comparable to each other.  

\begin{figure}[htbp]
\begin{center}
\includegraphics[scale=0.8]{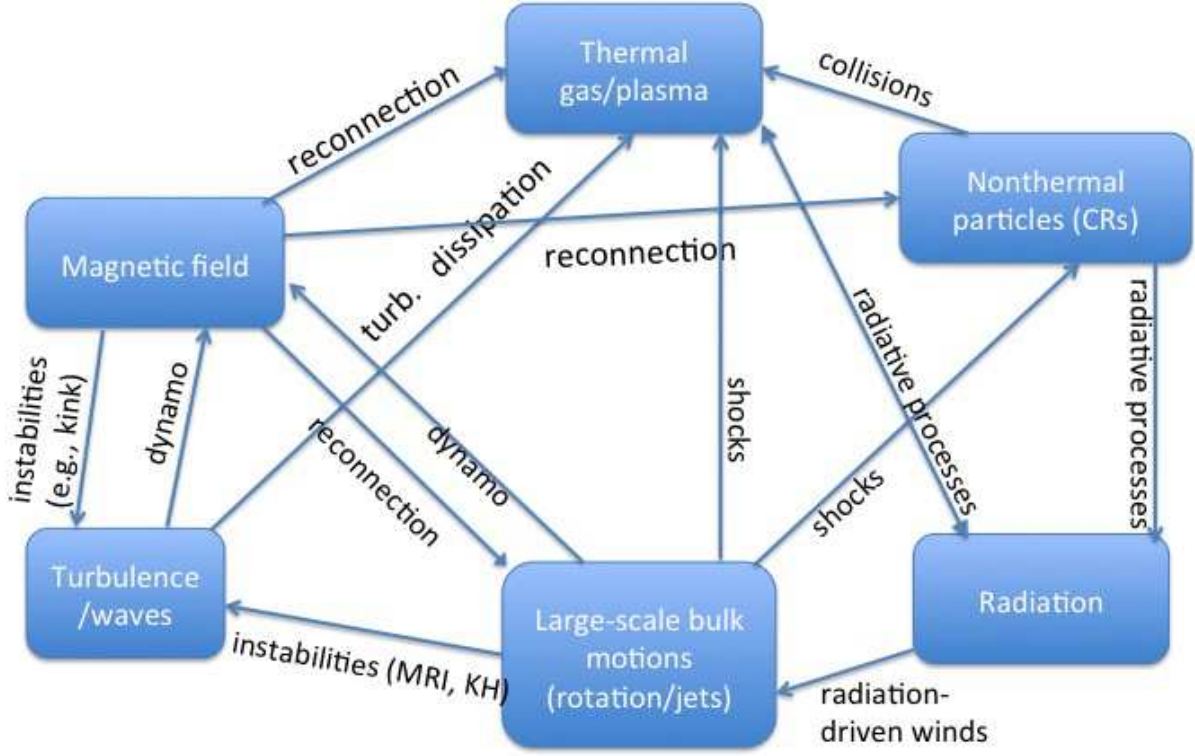}
\caption{{\bf Constituents and Energy Exchange Processes in Complex Cosmic Plasmas. 
}}
\label{fig-1}
\end{center}
\end{figure}

The plasma constituents exchange energy with each other through various physical {\it processes}, as illustrated in figure~\ref{fig-1}.  For example, the process by which bulk kinetic energy is converted into magnetic energy is called the dynamo; the process that converts magnetic energy into the thermal, bulk kinetic, and nonthermal particle energies is magnetic reconnection; the process of conversion of large-scale bulk kinetic energy to heat and nonthermal particles is a shock. 
Other important energy-exchange processes include radiative processes (converting particle kinetic energy to radiation and vice versa) and turbulent dissipation.  Some of these processes involve instabilities, for example:   the Rayleigh-Taylor (RT) instability converts gravitational energy to bulk kinetic; the magneto-rotational instability (MRI) converts the energy associated with a large-scale differential rotation to small-scale kinetic and magnetic energy; the Kelvin-Helmholtz (KH) instability converts large-scale kinetic energy to turbulent kinetic and magnetic energy; the kink instability converts magnetic energy to bulk kinetic, etc. 


Studying the energy conversion/exchange  processes between various constitients in complex astrophysical plasmas, i.e., of the various ways in which they interact with each other, is the main scope of {\it Plasma Astrophysics}.


\subsection{Extreme plasma astrophysics}
\label{subsec-extreme-astrophysics}

We shall now transition from the general discussion of plasma astrophysics given in the previous section to a more focused discussion of a special sub-area that we, in this review, will call the {\it Extreme Plasma Astrophysics}.  By this we shall mean plasma physics relevant to astrophysical environments with ultra-strong magnetic fields --- magnetic fields approaching and exceeding the critical quantum-electrodynamical (QED) magnetic field strength 
\beq
B_* \equiv  {{m_{\rm e}^2 c^3}\over{e\hbar }}  \simeq 4.4\times 10^{13}\ {\rm G} \, .
\eeq
This critical field has a clear physical interpretation that the corresponding electron cyclotron energy~$\hbar \Omega_e$ is equal to the electron rest-mass energy~$m_{\rm e} c^2 = 0.51\, {\rm MeV}$. We will call magnetic fields exceeding this value, "ultra-strong" or "super-critical" (see section~\ref{subsec-B_*-physics}).

However, since the general spirit of this review is to look at plasma astrophysics as a study of energy exchange processes, what we here are really interested in is, in general, astrophysical environments with energy densities corresponding to that of the critical field --- i.e., energy densities exceeding $10^{26}\, {\rm erg/cm}^3$; this also corresponds to blackbody radiation energy density ($aT^4$, where $a \simeq 7.57 \times 10^{-15}\, {\rm erg\, cm^{-3}\, K^{-4}}$ is the radiation constant) for a temperature of order the electron rest mass, $k_B T\sim m_{\rm e} c^2$. For the purposes of this review, we will call astrophysical environments with this level of energy densities "extreme astrophysical environments".  Our main focus will be on plasma-physical aspects of such ultra-magnetized, and often relativistically hot and radiation-dominated plasmas. 
The character of the energy exchange processes discussed in the previous section may be significantly altered in these environments. 

We would like to remark that, in addition to its direct astrophysical applications described below, this new frontier of plasma astrophysics is also interesting from the fundamental-physics point of view: it lies at the intersection of plasma physics, quantum field theory (namely, QED), high-energy-density physics, and high-energy astrophysics.
A rich variety of exotic physical effects come into play in these extreme plasmas, making this 
inter-disciplinary new direction intellectually challenging and attractive. 

This review is structured as follows.
In section~\ref{sec-extreme-astro-environments}  we will describe various astrophysical contexts in which extreme, ultra-magnetized plasma environments are encountered: 
(1) interiors of neutron stars (section~\ref{subsec-NS-interior}); 
(2) magnetospheres of magnetars (section~\ref{subsec-magnetars});
(3) central engines of supernovae and gamma-ray bursts (section~\ref{subsec-SN_GRB}).
The basic theoretical framework for various models of extreme plasmas is described in 
section~\ref{sec-Quantum-Plasma-Physics}.  
This section is primarily devoted to an exposition of quantum plasma physics.
In particular, in section~\ref{subsec-B_*-physics} we discuss the basic physics of super-critical magnetic fields. 
In section~\ref{subsec-nonrel-QP}  we discuss non-relativistic quantum plasma physics, including the fundamental theoretical formalism for quantum kinetic theory (section~\ref{subsubsec-Wigner-Moyal}), its application to linear waves and instabilities (section~\ref{subsubsec-quantum-waves}), quantum fluid theory (section~\ref{subsubsec-nonrel-quantum-fluid}), and spin quantum plasmas (section~\ref{subsubsec-nonrel-spin-plasma}).  In section~\ref{subsec-rel-QP} we describe the recent advances in relativistic quantum plasma physics, namely, 
for spinless Klein-Gordon plasmas (section~\ref{subsubsec-KG}), for spin-1/2 Dirac fermion plasmas (section~\ref{subsubsec-Dirac}), and finally, for relativistic quantum hydrodynamics (section~\ref{subsubsec-rel-hydro}). 
In section~\ref{sec-astro-applications} we turn to astrophysical applications of QED plasma physics and discuss several examples: 
propagation of high-energy radiation through magnetar magnetospheres (section~\ref{subsec-light-propagation}), 
large-scale MHD processes (section~\ref{subsec-MHD}), 
thermodynamics of extreme astrophysical environments (section~\ref{subsec-thermo}), 
interaction between magnetospheric currents and a magnetar surface (section~\ref{subsec-surface}),
and reconnection of ultra-strong magnetic fields in the context of magnetar flares and GRB central engines (section~\ref{subsec-reconnection}). 
A summary of the review and concluding remarks are presented in section~\ref{sec-summary}.


\section{Systems with extreme astrophysical environments}
\label{sec-extreme-astro-environments}

There are several important classes of high-energy astrophysical systems where the extreme ultra-magnetized conditions discussed in the previous section are encountered: 
\begin{enumerate}
\item{Interiors of neutron stars;}
\item{Magnetospheres of magnetars;} 
\item{Central Engines of Supernovae (SNe) and Gamma-Ray Bursts (GRBs) and inner parts of GRB jets.}
\end{enumerate}
This section is devoted to an overview of these systems with a special emphasis on outstanding plasma-physical problems relevant to them.


 \subsection{Magnetic fields in the interiors of neutron stars}
 \label{subsec-NS-interior}
 
Neutron stars (NSs) are one of the three types, along with white dwarfs (WDs) and black holes (BHs),  of compact objects that remain in the aftermath of the death of main-sequence  stars (see \cite{Shapiro_Teukolsky-1983-book} for review).  Like BHs, NSs are born in SN explosions marking the death of massive stars [i.e., stars with initial Zero-Age Main Sequence (ZAMS) masses greater than roughly 8 solar masses]. 

Neutron stars typically have masses between 1 and 2 solar masses, with most of the masses clustered around the Chandrasekhar mass of $M_{\rm Ch} \simeq 1.4 \, M_\odot$, where $M_\odot \simeq 2\times 10^{33}\, {\rm g}$ is the solar mass.  The most striking characteristics of NSs are perhaps their very small radii, of order 10~km,  just a few times larger than the corresponding Schwarzschild radius, and, correspondingly, their extremely high central densities, comparable to the density of the nuclear matter, $\rho_{\rm nuc} \simeq 3\times 10^{14}\, {\rm g \, cm}^{-3}$.  In essence, the interior of a neutron star can be seen as just one giant nucleus. As the NS name suggests, the composition of the interior is dominated by neutrons; they comprise more than 90\% of NS mass and their degeneracy pressure supports the star against gravitational collapse. In addition, there is a small admixture (a few per cent) of protons and electrons.  NSs are essentially macroscopic quantum objects: their interiors are believed to be superconducting and superfluid.  Another unique feature that distinguishes NSs from all other stellar-mass celestial objects is that, with the exception of newly-born NSs, they are believed to be covered with a solid crust, about a kilometer thick, composed mainly of iron with a density of order $10^6\, {\rm g\, cm}^{-3}$. 

Since NSs represent the end-state of stellar evolution for a substantial fraction of the stellar population 
(namely, stars with ZAMS masses between 8 and about 20-25~$M_\odot$), there should be literally billions of old NSs flying around in our Galaxy and its halo.%
\footnote{NSs are usually born with large so-called "natal kicks" imparted to them during the SN explosion  
and have typical velocities of hundreds of km/sec, comparable to the virial velocity in the Galaxy. These velocities are much higher than the velocity dispersion of normal stars in the galactic disk, so NSs are expected to populate a torus that is much thicker than the stellar disk.}
However, because they are so compact, they are very dim in the optical and so are very difficult to detect.  Observationally, the main classes of neutron stars are: \\ 
(1)  classical radio-pulsars, the first kind of NSs discovered, usually found by their pulsed periodic radio emission; currently there are almost 2000 of them known; \\
(2) accreting NSs in Galactic X-ray binaries (XRBs), usually discovered as powerful X-ray sources (about 5\% of the total known NS population), including X-ray pulsars and X-ray bursters; \\
(3) magnetars --- NSs with magnetic fields of $10^{14}-10^{15}\, {\rm G}$, identified observationally as anomalous X-ray pulsars (AXPs) and soft gamma repeaters (SGRs), see section~\ref{subsec-magnetars}.  There are just a few known magnetars in our Galaxy and the Magellanic clouds, but they are believed to comprise about 10\% of the entire NS population at birth, e.g., \cite{Kulkarni_Frail-1993, Kouveliotou_etal-1994, Kouveliotou_etal-1998}.

An outstanding plasma astrophysics problem in the context of NSs has been the origin and evolution of their magnetic fields.  It is remarkable that, despite the very narrow spread of NS masses and sizes, they seem to  exhibit a huge spread in their (dipole) magnetic field strengths.  The typical magnetic fields inferred in most regular NSs, e.g., radio- and X-ray pulsars, are of the order of $10^{12}$~G, but there are also a large number of NSs with much weaker fields, perhaps as low as $10^9$~G and lower, in old systems such as recycled millisecond radio-pulsars and type-I X-ray Bursters in Low-Mass XRBs (LMXBs).  On the strong-field side, there are magnetars --- NS with estimated magnetic fields of order $10^{15}$~G.  Explaining this great diversity of NS magnetic fields is an important problem in modern astrophysics.  

Some of this diversity may reflect a possible wide range of the field strengths that NSs are born with, but some probably may be attributed to magnetic field decay later in the NS's life.  It is important to understand the role of both of these stages.  
Thus, the overall problem of explaining magnetic field distribution in NSs naturally splits into two rather independent sub-problems. The first one is the problem of establishing the initial distribution of NSs with respect to their magnetic fields, i.e., the initial NS magnetic field function~\cite{Spruit-2008}.  The second problem is that of long-term evolution of the magnetic field in old NSs.

The first problem can be called the NS magnetogenesis problem: it aims at understanding the processes generating the initial magnetic field in NSs just as they are born in a core-collapse SN explosion.  This problem is inextricably related to the problem of understanding the mechanisms of SN explosions (see section~\ref{subsec-SN_GRB}).  
There is a large number of astrophysically important questions that one would like to address: 
Why are some (perhaps 10 percent) neutron stars (namely, magnetars) endowed with ultra-strong ($10^{14}-10^{15}$~G) magnetic fields at birth, whereas most NSs have much weaker ($10^{12}$~G) fields? 
Are there NSs with very weak initial magnetic fields, e.g., of order $10^9-10^{10}$~G? 
What basic parameters of the progenitor massive star (e.g., its mass, rotation rate, metallicity, etc.) influence or control the large-scale magnetic field with which a newly born NS emerges after the SN explosion clears? 
How strong is this influence, i.e., is the resulting initial magnetic field essentially predetermined by the above progenitor's parameters or is it largely a random quantity?
Is there any connection between NS fields and natal kicks? 

The NS magnetogenesis problem is very difficult to address observationally --- we simply do not have observational tools to measure the initial NS magnetic field distribution function.  However, there have been a significant number of theoretical papers devoted to this subject in the past couple of decades, in particular, in relation to the origin of magnetars.  It is generally believed, to a large degree by analogy with the solar magnetic field, that a NS's magnetic field is produced by a large-scale $\alpha\,\Omega$ MHD dynamo in a rapidly rotating convective proto-NS in the first few seconds of its evolution~\cite{Duncan_Thompson-1992, Thompson_Duncan-1993, Spruit-2008}.  Turbulent dynamo action is usually thought to require two essential ingredients: turbulence (driven by, e.g.,  thermal convection or by some other  instability) and a large-scale (differential) rotation. 
Both of these conditions are likely to be present in young proto-NSs. 
First, in the first few seconds of their evolution, NSs are indeed believed to experience vigorous convection driven by neutrino cooling (e.g., \cite{Burrows_Lattimer-1988}). In addition, they may become turbulent as a result of some active MHD instability, such as the MRI \cite{Akiyama_etal-2003, Ardeljan_etal-2005, Moiseenko_etal-2006} or the Tayler-Spruit process \cite{Tayler-1973, Spruit-1999, Spruit-2002}. 
As for rotation --- the second important ingredient needed for dynamo --- proto-NSs can indeed be rapid rotators: since the gravitational collapse of a WD-like progenitor core is very rapid (a couple hundred milliseconds), angular momentum is conserved during this process and so the emerging NS rotation rate is basically determined by the rotation rate of the core.  For example, as was suggested by Duncan and Thompson~\cite{Duncan_Thompson-1992}, for a newly-born NS to develop a large-scale magnetic field of order $10^{15}$~G, it needs to have an initial rotation period of order a millisecond.
\footnote{The slow (a few seconds) spin periods currently observed in AXPs and SGRs are explained by the very rapid spin-down soon after their birth.} 
This in turn requires, by angular momentum conservation, the progenitor core to have a rotation period of just a few seconds. 
This argument illustrates an important link that ties the NS rotation, and hence, presumably, its large-scale magnetic field and its potential as a GRB central engine (see section~\ref{subsec-SN_GRB}), to the parameters of the progenitor star. 
As was suggested by Heger {\it et al} \cite{Heger_etal-2005}, such a rapid progenitor core rotation rate is actually very difficult to maintain during the preceding evolution phase of the progenitor, due to angular momentum losses to the outer stellar envelope. This implies that the birth of a NS with a millisecond spin period must be a rare event. 
[Note that there is a class of NSs that have millisecond periods (so-called recycled millisecond pulsars, see below) but these are probably not their initial periods --- these are old, weak magnetic field NSs in binary systems that have been spun up by accretion.]  

One important factor that hampers our understanding of magnetic field generation in proto-NSs is that we still do not have a reliable predictive theory of MHD dynamo.  While significant progress in reproducing key features of the solar dynamo has been achieved with the help of numerical simulations in the past few years (e.g., \cite{Brown_etal-2010, Ghizaru_etal-2010}), this success is yet to be extended  to the NS magnetogenesis problem.  Thus, the question of how the neutrino-convection-driven large-scale MHD dynamo operates at extreme densities and magnetic fields in a rapidly cooling (by neutrino emission) hot proto-NS during the first few seconds of its life remains an open frontier of NS research.

The second important aspect of the NS magnetic field distribution problem is the problem of long-term  evolution of the magnetic field, in particular, their decay over time.  Two important astrophysical contexts in which this problem arises are: (1) the decay of magnetic field in accreting NSs in LMXBs and the origin of recycled millisecond pulsars, and (2) the decay of magnetic field in magnetars on $10^2-10^4$ year time scale and the resulting cessation of magnetar activity. 

Long-term magnetic field decay in normal isolated NSs due to processes such as ohmic dissipation, ambipolar diffusion, and Hall drift, and its effect on the NS thermal evolution has been considered in a number of studies \cite{Haensel_etal-1990, Romani-1990, Goldreich_Reisenegger-1992, Shalybkov_Urpin-1997, Vainshtein_etal-2000, Geppert_Rheinhardt-2002, Cumming_etal-2004, Pons_Geppert-2007}. 
In addition, the problem of magnetic field decay/burial in accreting NSs has attracted a lot of attention and has been the subject of a significant research effort in recent years due to its importance for understanding the origin of  recycled millisecond radio-pulsars (see \cite{Bhattacharya-1995, Konar-2013} for reviews).  These are relatively old NSs found in some binary systems, which have presumably been spun-up to millisecond periods by accretion sometime in the past.  Maintaining such a rapid rotation rate over an extended period of time, however, implies that the standard magnetospheric spindown mechanism is very weak in these systems. In particular, the observed very slow spin-down rates indicate large-scale dipolar magnetic fields as low as $10^8-10^9$~G,  significantly lower than the standard $10^{12}$~G fields of "normal" isolated pulsars.  One popular explanation for this is an effective "burial" of the NS magnetic field during the accretion stage (e.g., \cite{Romani-1990, Brown_Bildsten-1998, Cumming_etal-2001}). This is consistent with the observation that older actively accreting NSs in LMXBs typically have weaker inferred magnetic fields ($10^9-10^{10}$~G) than younger accreting NSs in HMXB, detected as  X-ray pulsars ($\sim 10^{12}$~G).

In the context of magnetars, magnetic field decay on the time scale of $10^4$ years and its effect on the star's thermal evolution and on its potential for powering the X-ray emission in Anomalous X-ray Pulsars (AXPs) has been investigated by several authors, e.g.,  \cite{Thompson_Duncan-1996, Heyl_Hernquist-1997c, Heyl_Kulkarni-1998, Colpi_etal-2000, Arras_etal-2004, Pons_Geppert-2007, Aguilera_etal-2008, Giannios-2010, Cooper_Kaplan-2010, Dall'Osso_etal-2012}.  We will discuss some of these processes in section~\ref{subsec-magnetars}.

Finally, there is a question of the structure of the magnetic field inside NSs, including magnetars, once they are fully formed.  Since NSs stop being convective fairly soon, just a few seconds after their birth, the usual convective MHD dynamo processes no longer operate in such NSs. Differential rotation also decays quickly and hence the Tayler-Spruit \cite{Tayler-1973, Spruit-2002} dynamo action also ceases. 
The magnetic field inside a NS then eventually relaxes to a stable lowest-energy magnetostatic equilibrium configuration, consistent with global magnetic helicity constraints and with field-line pinning in the solid crust.   There has been a significant amount of theoretical work in recent years on calculating such equilibrium configurations and assessing their stability \cite{Braithwaite_Spruit-2006, Geppert_Rheinhardt-2006, Kiuchi_Kotake-2008, Broderick_Narayan-2008, Braithwaite-2009, Reisenegger-2009, Ciolfi_etal-2009, Ciolfi_etal-2010, Gourgouliatos_etal-2013}.


 \subsection{Magnetospheres of magnetars}
 \label{subsec-magnetars}
 
Magnetars are isolated young neutron stars with large-scale magnetic fields of order~$10^{15}\, {\rm G}$ --- much stronger than the "mere" $\sim 10^{12}$~G fields typical for normal NSs (e.g., radio-pulsars or X-ray pulsars) and also stronger than the critical quantum magnetic field~$B_*\sim 4 \times 10^{13}\, {\rm G}$.  
Since the focus of this review is on physical processes, rather than on observational phenomenology, it will be convenient for us to {\it define} magnetars as a class of neutron stars with large-scale magnetic fields in excess of~$B_*$. 

There is a number of  interesting phenomena exhibited by magnetars, and  understanding them often requires a certain  degree of plasma physics.  Observationally, magnetars are detected through their powerful X-ray and $\gamma$-ray emission (see, e.g., \cite{Mereghetti-2008} for a review). 
Based on their high-energy behaviour, magnetars are usually subdivided into two classes --- Anomalous X-ray Pulsars (AXPs) and soft gamma repeaters (SGRs).  There are only a handful of objects of each class known. 
Both AXPs and SGRs are characterized by a powerful persistent X-ray emission ($L_X = 10^{34}-10^{36}$ erg/s),  including a nonthermal power-law hard X-ray component observed in some AXPs (extending to at least 100 keV). However, in addition to this X-ray emission, SGRs also produce occasional very intense outbursts of gamma-ray activity (see below). 

Both classes of magnetars have relatively slow rotation $P= 5-12$~sec, detected through the modulation of their persistent X-ray emission, but relatively high spin-down rates ($\dot{P} \sim 10^{-10} - 10^{-12}$) and hence young spin-down ages ($P/\dot{P} \lesssim 10^3 - 10^5$ years, consistent with the ages of their associated SN remnants). This provides a key evidence for high magnetic fields in these objects.  SGRs have somewhat higher spin-down rates than AXPs and hence are believed to be younger ($10^3$ years instead of $10^4-10^5$ for~AXPs).  This has led to the suggestion that after a few thousand years SGRs lose their $\gamma$-ray activity and evolve into~AXPs. 
In any case, an important implication of the very young ages of magnetars is that, even though the number of known magnetars is very small compared with the number of known regular NSs such as radio-pulsars (which are, on average, much older, with typical spin-down ages of millions of years), a rather significant fraction of all NSs, perhaps as high as $\sim 10$\%, are born as magnetars  \cite{Kulkarni_Frail-1993, Kouveliotou_etal-1994, Kouveliotou_etal-1998}.


\subsubsection{Persistent X-ray emission of AXPs and SGRs.}
\label{subsubsec-magnetars-Xray}

The X-ray emission of both AXPs and SGRs is so powerful that explaining it becomes a nontrivial task and leads to an additional, independent line of evidence for the presence of very strong magnetic fields in these objects.  Indeed, the observed X-ray power ($10^{34}-10^{36} \, {\rm erg/s}$) cannot be attributed to accretion because these systems do not show any signs of having binary companions. Furthermore, it is so strong that it cannot be attributed to the NS spin-down power, because magnetars are relatively slow rotators and their spin-down power is small (of order $10^{32} \, {\rm erg/s}$).  Instead, the emission is believed to be powered by the dissipation of their huge magnetic energy \cite{Thompson_Duncan-1995, Thompson_Duncan-1996, Kouveliotou_etal-1999, Woods_Thompson-2006}.  Indeed, one can easily convince herself that the magnetic energy of a magnetar ($E_{\rm mag} \sim B^2 R_{\rm NS}^3/6$) provides an adequate reservoir of energy to power the high-energy (X-ray and gamma-ray) emission for $\sim 10^4$ years. 

In a generally accepted scenario developed in a series of papers by Thompson and Duncan \cite{Duncan_Thompson-1992, Thompson_Duncan-1995, Thompson_Duncan-1996, Thompson_Duncan-2001, Thompson_etal-2002}, the ultimate energy source powering magnetar high-energy emission is the free magnetic energy of the highly twisted, entangled magnetic field inside the~NS.  From time to time this internal field experiences episodes of relaxation, in which its twist is transmitted to the external segments of the magnetic field lines that protrude above the star's surface, i.e., that  constitute the star's magnetosphere.  Thus, this  process involves sheared motions of the footpoints of the external field lines, which has to happen despite the fact that neutron stars (including magnetars) are covered by a solid rigid crust into which the magnetospheric field lines are effectively frozen.  The reason why this may be possible at all is that the stresses of the twisted magnetic field in a magnetar are strong enough to either break the crust in discrete episodes, called star-quakes, or to lead to gradual plastic deformations of the crust.  In either case, the field lines in the magnetosphere get twisted and the field becomes a non-potential, force-free field.  Correspondingly, the magnetic energy contained in the magnetosphere increases relative to that of the potential magnetic field.  It is the dissipation of this excess magnetic energy that subsequently powers both the persistent X-ray emission in AXPs and SGRs and also the occasional $\gamma$-ray flares in~SGRs.  

An important mechanism for the X-ray emission involves the interaction of the electric currents in the twisted magnetosphere with the surface of the neutron star \cite{Beloborodov_Thompson-2007}. 
This interaction results from an intense beam of current-carrying particles striking the surface of the NS and causing strong localized heating of the surface to hard-X-ray temperatures (100~keV). 
In particular, the currents with densities $j \sim c B/4\pi R_{\rm NS}$, which have to flow in the force-free magnetosphere in order to support finite twist angles of order 1 rad, are so intense that they require a high number density of charge carriers, of order $n\sim j/ec \sim B/(4\pi e R_{\rm NS}) \sim 10^{17}\, {\rm cm}^{-3}$ \cite{Thompson_etal-2002}.  These particles  must be drawn out from the NS surface or created above it via a pair cascade discharge, similar to what happens in regular pulsars \cite{Arons_Scharlemann-1979, Arons-1979}.  
The kinetic plasma-physical processes  involved in these mechanisms in the context of magnetars have been discussed briefly by Beloborodov and Thompson \cite{Thompson_Beloborodov-2005, Beloborodov_Thompson-2007} (see also \cite{Istomin_Sobyanin-2007}). 
In general, however, microscopic plasma phenomena in the presence of $\sim B_*$ fields can be modified by quantum-electrodynamic effects in nontrivial ways.  The ramifications of these effects for astrophysical processes, such as pair production in active magnetar magnetospheres, are not yet understood in full detail, but some progress in assessing quantum-relativistic effects on a few  simple, classically well-understood basic kinetic plasma phenomena has been made, see section~\ref{subsec-rel-QP}.
 
Whatever the mechanism for populating the magnetospheric currents with plasma is, at the end of the day these currents are basically intense particle beams flowing along the field lines from one foot-point to the other.  As these beams strike the NS surface, they are suddenly stopped and their energy is dissipated, causing very intense local heating at the foot-points of twisted magnetic field lines.  The resulting (effectively resistive) surface dissipation of the magnetospheric currents thus powers the persistent X-ray emission of magnetars, somewhat similar to the processes powering the surface emission in two-ribbon solar flares.  Ultimately, the energy comes from the magnetic energy of the nonpotential magnetic field, because the resistive electric field associated with the current dissipation at the footpoints implies marching of the field lines across the NS surface in the direction of decreasing twist, which takes the field closer to the potential state and hence reduces the magnetic energy. 

The questions of how exactly the intense current beams are stopped when they hit the magnetar surface and how the kinetic energy of the particles is dissipated and distributed in the surrounding dense material motivate a whole set of interesting plasma-physical problems.  In particular, it is possible that the impinging beams excite kinetic micro-instabilities such as the Buneman and ion-acoustic instabilities. The nonlinear development of these instabilities quickly leads to small-scale Langmuir turbulence;  the effective anomalous resistivity associated with this turbulence is likely to play a dominant (relative to the particle-particle Coulomb collisions) role in stopping the beam and dissipating its current \cite{Beloborodov_Thompson-2007}.  But how exactly this turbulence develops, what controls the resulting level of the effective resistivity and hence the corresponding stopping length of the beam, and how deep the energy-deposition region lies, is still rather poorly understood, especially in the presence of the ultra-strong magnetic field. These questions provide a strong astrophysical motivation for studying waves and instabilities in ultra-magnetized relativistic quantum plasmas, a subject that we will discuss in more detail in section~\ref{sec-Quantum-Plasma-Physics}.

In addition to the NS surface, another source of resistivity is the resonant cyclotron and Compton scattering of the X-ray photons emitted from the star's surface by the current-carrying electrons (and/or positrons) high above it, at distances of several tens of kilometers \cite{Thompson_Beloborodov-2005}. 
As was discussed by Thompson {\it et al} \cite{Thompson_etal-2002}, the number density of charge carriers required to carry the necessary force-free currents in a strongly twisted magnetosphere (with twist angles of order 1 radian) is so high that the optical depth to resonant cyclotron scattering is automatically of order~1. This means that some of the photons bounce in the magnetosphere perhaps several times before they can escape, which  significantly alters the emerging X-ray spectrum \cite{Lyutikov_Gavriil-2006, Rea_etal-2008}. 
In particular, this multiple scattering can lead to the formation of a power law, similar to the way hard-X-ray power-law spectra are believed to be produced \cite{Sunyaev_Titarchuk-1980} in Comptonizing coronae of accreting black holes with a Compton $y$-parameter of order unity. 
This provides a promising explanation for the observed X-ray spectra of the persistent emission in active magnetars. The marked difference between the power-law indices in AXPs (-4) and SGRs (-2) can then be attributed to the differences in their Compton $y$ parameters.


\subsubsection{SGR giant $\gamma$-ray flares.}
\label{subsubsec-SGR-flares}

Most of the time, magnetars are found in a quiescent state, but a subclass of magnetars ---  namely, Soft Gamma Repeaters (SGRs) --- sometimes produce intense bursts of soft gamma-rays. The most dramatic manifestation of magnetar activity is the SGR giant flares --- the most intense Galactic events, releasing $10^{44}-10^{46}$~erg in just a fraction of a second, (see, e.g., \cite{Mazets_etal-1999, Palmer_etal-2005, Hurley_etal-2005}).  These flares have an interesting emission morphology consisting of several distinct components: (1)  a hard ($k_B T\sim 200$~keV) main {\it spike} lasting $\sim 0.25$~sec; (2) a long-lasting (hundreds of seconds) and softer ($k_B T\sim 20$~keV) {\it tail}  emission modulated by the stellar rotation, which is seen as a signature of a modest-size fireball confined close to the star and corotating with it; and (3) an even longer lasting {\it afterglow}. 

It is generally believed that the underlying mechanism for powering these flares is, again, the release of the free energy of a non-potential magnetic field in the neutron star's magnetosphere, twisted up by the star's crustal displacements~\cite{Thompson_Duncan-1995, Thompson_Duncan-2001, Thompson_etal-2002}. 
However, the specific mechanism for magnetic energy dissipation is not yet firmly established. 
It has been suggested, on general grounds and by analogy with solar flares, that the mechanism powering SGR flares is magnetic reconnection~\cite{Thompson_Duncan-1995, Thompson_Duncan-2001, Lyutikov-2003b, Lyutikov-2006a, Beloborodov-2009, Masada_etal-2010, Uzdensky-2011, Yu-2012, Parfrey_etal-2013, Yu_Huang-2013}. 
Indeed, observationally inferred rapid variation of magnetospheric properties, e.g., of the spindown rate, indicates that an SGR  magnetosphere is active, similar to the solar corona. 
This generally implies that the field-line footpoint motions gradually twist up and stress the field, raising the free magnetic energy, as discussed above.  This process, however, cannot last indefinitely.  
Any such evolution in a semi-unbounded magnetosphere generically leads to the inflation of some of the field lines (e.g., \cite{Aly-1985}). 
At some point, the twist imparted to the  field lines by the crustal motions exceeds a critical value and the stressed force-free magnetosphere effectively opens up in an explosive manner and then either becomes unstable or undergoes a loss of equilibrium (see \cite{Uzdensky-2002b} and references therein).  
Either outcome leads to the formation of thin current layers~\cite{Uzdensky-2002a}. 
Just like in the solar corona or in the Earth's magnetosphere, this sets up the conditions for the onset of magnetic reconnection.  Similar large-scale MHD processes associated with the evolution of force-free magnetic fields under sheared footpoint motions are believed to be happening also in force-free coronae of the Sun and other active stars and in the coronae of astrophysical  accretion disks, e.g., around young stars and NSs \cite{Uzdensky-2004}.  These processes are very complex, especially in 3D, and are not yet completely understood even in the more familiar solar context, despite very extensive research over the past several decades. 
However, in the case of a magnetar magnetosphere, a number of simple illustrative examples of an axisymmetric over-twisted magnetosphere, in which a current sheet forms at the equator, have been explored  \cite{Thompson_etal-2002, Lyutikov-2006a,  Beloborodov-2009, Yu-2012, Parfrey_etal-2012, Parfrey_etal-2013, Lyutikov-2013}, including numerical relativistic force-free models \cite{Parfrey_etal-2012, Parfrey_etal-2013}.  Furthermore, it has been argued that the transient opening of the magnetosphere during a flare also leads to other important observable signatures such as a rapid spindown (an "anti-glitch") of the magnetar \cite{Parfrey_etal-2012, Lyutikov-2013}.

Invoking the same solar-flare analogy, one can conjecture that a magnetar reconnection event produces two oppositely directed outflows carrying the dissipated energy (in roughly equal amounts, typically of order $10^{44}$~erg) out of the reconnection region \cite{Masada_etal-2010, Uzdensky-2011}. 
One outflow is directed towards the NS and forms a hot fireball confined by the strong post-flare magnetic loops next to the star. The other forms a hot plasma blob ejected relativistically away from the star. 
Even though they may start out similar, the fates and observational appearances of these two fireballs are very different. 
The freely expanding fireball becomes optically thin and produces the main hard (200~keV) spike lasting a fraction of a second and accounting for most of the observed flare energy; later, it may also power an extended afterglow at much larger distances \cite{Nakar_etal-2005}.  In contrast, the fireball magnetically confined  next to the star gradually cools by radiation and is responsible for the long-lasting (hundreds of seconds) and relatively cool (20 keV) tail emission modulated by the star's rotation. 

We would like to note that, whereas on general grounds there seems to be a good case for the reconnection scenario of magnetar flares, how exactly this happens in such an extreme  environment, and what its consequences are, has not yet been systematically explored and is still poorly understood. 
It is therefore important to investigate the underlying relativistic quantum plasma physics (see section~\ref{sec-Quantum-Plasma-Physics}) of reconnection of magnetar-strength fields and to connect the theory with the key observational constraints: the short (milliseconds) flare rise time, and the durations and the temperatures of the main $\gamma$-ray spike and of the tail emission. 
Magnetic reconnection of magnetar-strength fields  is the most dramatic and extreme example of  astrophysical reconnection; developing a detailed quantitative picture of this process is an important problem in plasma astrophysics. We will discuss this problem in more detail in section~\ref{subsec-reconnection}.


\subsubsection{Propagation of electromagnetic radiation through a magnetar magnetosphere.}

It has to be noted that, ultimately, our only observational window into what happens in magnetar systems lies in the high-energy radiation that we detect at Earth.  For this reason, our interpretation of magnetar phenomena critically depends on a good understanding of the propagation of the high-energy radiation through the magnetosphere.  Whatever light is emitted from the inner part of the magnetosphere, e.g., from the footpoints of current-carrying twisted magnetic loops, it inevitably has to make its way through a region of ultra-strong magnetic field in order to get to us.  The magnetic field strongly interacts with high-energy photons and greatly affects their propagation via a number of QED effects, such as 1-photon pair creation, photon splitting, etc. (see \cite{Harding_Lai-2006} for review).  Furthermore, the presence of an ultra-strong magnetic field  alters the ways in which high-energy photons interact with the magnetospheric plasma, e.g., by modifying the Compton scattering cross-section. 
These processes are reviewed by Harding and Lai \cite{Harding_Lai-2006} and we will discuss them in section~\ref{subsec-light-propagation}. 


To sum up, important plasma physics problems pertaining to magnetar magnetospheres include:

{\bf (1)}  Kinetic plasma processes (including pair cascades and effective anomalous resistivity due to Buneman and ion-acoustic microturbulence) at the magnetosphere-surface interface --- the suggested place of dissipation of intense field-aligned currents in the twisted force-free magnetar magnetosphere, which leads to a strong localized surface heating that powers the persistent X-ray emission in AXPs and SGRs. 

{\bf (2)} Large-scale MHD evolution of the force-free magnetosphere (perhaps analogous to the solar corona), including  twisting and consequent inflation of the magnetic flux loops, leading to   field-line opening, loss of equilibrium,  MHD instabilities, current sheet formation, and reconnection onset. 

{\bf (3)} Reconnection in magnetar magnetospheres as a possible mechanism of SGR flares. 

{\bf (4)} Effects of ultra-strong magnetic field on photon propagation in a magnetar magnetosphere.


\subsection{Central engines of supernovae and gamma-ray bursts} 
\label{subsec-SN_GRB}

Another important example of extreme astrophysical environments is central engines and inner parts of magnetically-driven jets in supernovae (SNe) and Gamma-Ray Bursts (GRBs). 

Both SNe and GRBs are very important and still not fully understood celestial phenomena. 
These spectacular cosmic explosions, marking the death of massive stars, have fascinated both astrophysicists and the general public alike for decades. Trying to understand their progenitors and the explosion mechanisms has been the subject of extensive research efforts. Still, however, there is no complete consensus among scientists on some of the key principal questions, such as the actual explosion mechanism  or the nature of the central engine (for GRBs). These systems are very complex, rich in physics that is quite different from that encountered under any other circumstances.  Of relevance to our review, among the various uncertainties impeding our progress, there are many that have to do with extreme plasma physics, as we shall describe in this subsection. 

Before we proceed, we note that, although many basic physical characteristics of magnetar magnetospheres and SN/GRB central engines are similar (sizes, time scales, typical magnetic field strengths, etc.), 
there are also many important differences. 
The physical conditions in SN and GRB central engines are quite exceptional and are not encountered anywhere else in the Universe since the Big Bang.  In particular, the temperatures, reaching $10^9$ K, and the matter densities, of the order of the nuclear density of $\sim 10^{14}\, {\rm g\, cm^{-3}}$, and hence the plasma  pressures, are much higher here than around magnetars.   
Under these extreme conditions, the nuclear composition undergoes rapid changes as the material  transforms from predominantly iron in the progenitor core to neutron-rich matter of the resulting neutron star.  Also, neutrinos play a very important role in both the dynamics (exerting pressure) and, especially, the energetics of these explosions. 

Before we turn to the discussion of the  plasma-physical problems important to the operation of SN and GRB central engines, we shall give a brief basic introduction to each of them. 


\subsubsection{Supernovae.}
\label{subsubsec-SN}

SNe are some of the most energetic and spectacular events in the universe. 
Energetically, these stellar explosions are  second perhaps only to GRBs, but are much more numerous. 
In addition to being interesting by themselves,  SNe are also very important because of their profound influence on the galactic evolution. They are responsible for:  
heavy-element nucleosynthesis; 
dispersion of metals, from which dust is subsequently formed, into the ISM; 
stirring interstellar turbulence and galactic magnetogenesis; 
and cosmic-ray acceleration up to $10^{15}$~eV in SN shocks. 
They also represent a potential source of gravitational waves and are used as a standard candle in cosmology (Type Ia SNe). 

SNe basically come in two classes:  (1) Type Ia SNe, involving a thermonuclear explosion of a (carbon-oxygen) white dwarf (WD) reaching the Chandrasekhar limit; and (2) core-collapse SNe (Types Ibc and Type II), involving gravitational collapse of WD-like iron cores of massive ($\gtrsim 8\, M_\odot$) stars; the latter results in the formation of a~NS. The  development and outward propagation of a bounce shock then disrupts the rest of the star and leads to the ejection of the stellar envelope (see \cite{Woosley_Weaver-1986, Bethe-1990, Woosley_etal-2002, Janka-2012, Burrows-2013} for reviews). 

In a core-collapse SN, an electron-degenerate iron core of a massive star exceeds its Chandrasekhar mass limit and undergoes a sudden gravitational collapse, forming a proto-NS within a few hundreds of milliseconds. Because the equation of state  of nuclear matter in the newly born NS is very stiff, the inner core bounces back, and a so-called bounced shock is launched outward into the in-falling material of the outer core. Careful analyses of this process have shown, however, that the shock looses steam and stalls within a fraction of a second when its radius is about 100~km.  It then has to be  somehow revived within about a second after bounce, if it is to disrupt the star and produce a successful SN explosion. 
Much of the SN research in the past few decades has focused on finding a suitable mechanism for the shock revival and a number of possibilities have been put forward  (see, e.g., \cite{Bethe_Wilson-1985, Bethe-1990, Woosley_Weaver-1995,  Woosley_etal-2002, Janka_etal-2007, Janka-2012, Burrows-2013}). 

At first, the problem does not appear to be that challenging from the overall energetics point of view.  Indeed, the energy needed for a SN explosion (just about $10^{51}$ ergs) is roughly 1 percent of the available gravitational energy released in the collapse of the progenitor core \cite{Colgate_White-1966}. However, the problem stems from the fact that almost all of this released energy is carried away by neutrinos \cite{Colgate_White-1966} as the newly-formed proto-NS cools down over several seconds. The detection, by the Kamiokande~II neutrino detector in Japan, of a definite neutrino signal associated with the 1987 SN in the Large Magellanic cloud \cite{Hirata_etal-1987}, in line with theoretical expectations, has provided us with a direct experimental confirmation of this picture.  Even though neutrinos are dominant in the overall core-collapse energy budget, it is not clear whether neutrino heating of the post-shock region around the proto-NS (e.g., via $\nu\bar{\nu}$  annihilation into electrons and positrons \cite{Goodman_etal-1987}) can, by itself, be sufficient to revive the shock and explode the star (e.g., \cite{Bethe_Wilson-1985, Bethe-1990, Janka_etal-2007}). The reason for this is that neutrinos do not couple well to stellar material and so imparting  even 1\% of their energy to the  matter is actually a nontrivial task.  The effectiveness of neutrino heating and its role in the SN explosion may be affected by the 2D and 3D turbulent motions in the post-shock region (\cite{Nordhaus_etal-2010}), perhaps driven by hydrodynamic instabilities such as the Standing Accretion Shock Instability (SASI) \cite{Blondin_etal-2003, Blondin_Mezzacappa-2006}.  Thus, modern numerical models suggest that whether or not one gets a successful explosion may depend on the simulation dimensionality.  It may well be that full 3D high-resolution neutrino-radiation-hydrodynamic numerical models are necessary to settle the question of whether the neutrino heating mechanism can, by itself, produce a SN explosion.  This direction is an active area of research and we can hope to expect some clarity within the next few years. 

In contrast to neutrinos, sound waves \cite{Burrows_etal-2006a, Burrows_etal-2007a} and magnetic fields couple to the gas quite well.  For this reason, even though they are energetically subdominant, both have been proposed to be the key element in SNe.  Since the main focus of this review is on plasma-physical aspects of extreme environments, here we will consider only the MHD (i.e., magneto-rotational) mechanisms, which are also relevant to GRBs (see below).  This subject is closely related to the problem of NS magnetogenesis, discussed in section~\ref{subsec-NS-interior}.

The idea that magnetic fields, together with a rapid NS rotation, may play an important role in powering a SN explosion is actually rather old.  It was proposed as a possible SN mechanism  just a few years after the discovery of radio-pulsars and the determination of their Tera-Gauss magnetic fields (e.g., \cite{LeBlanc_Wilson-1970, Ostriker_Gunn-1971, Bisnovatyi-Kogan-1970, Bisnovatyi-Kogan_etal-1976, Kundt-1976, Meier_etal-1976}). 
The main idea is that rapid differential rotation in the central engine can wrap up the poloidal (meridional) magnetic field to produce a toroidal field of order $10^{15}$~G, strong enough to affect the dynamics and even produce collimated outflows (which may help explain collimated jets in GRBs). 
The first 2D MHD axisymmetric simulations of this mechanism were conducted more than four decades ago by Le Blanc and Wilson \cite{LeBlanc_Wilson-1970}, who, however, used unrealistically large values of the initial magnetic field.  Later, versions of this mechanism were  investigated numerically by Symbalisty \cite{Symbalisty-1984} and then by other researchers (e.g, \cite{Ardelyan_etal-1987, Kotake_etal-2004, Yamada_Sawai-2004}).
Magneto-rotational effects on SN explosions were also investigated in detail by Wheeler {\it et al} \cite{Wheeler_etal-2000, Wheeler_etal-2002}. 
These models mostly focused on just the straight-forward effect of linear-in-time toroidal field generation by wrapping up the poloidal magnetic field lines by the rapid differential rotation that may exist between the NS and its surroundings. They were based on the simple estimate that a millisecond NS with a poloidal field of $10^{12}$~G will take just about a second to wrap the field up to a magnetar level ($10^{15}$~G), at which the field will start to exert a dynamic influence. 
However, several years after the re-discovery of the magneto-rotational instability (MRI) in the 1990s by Balbus and Hawley \cite{Balbus_Hawley-1991} and its successful application to the problem of angular momentum transport in accretion disks, similar ideas started to percolate into the SN research. 
In particular, Akiyama {\it et al} \cite{Akiyama_etal-2003} investigated the operation of MRI turbulence in the context of core-collapse SNe. They suggested that a weak seed magnetic field is greatly amplified by the resulting MRI dynamo action up to a saturation level of order $10^{15}-10^{16}$~G within the first 300 ms after bounce.  This field could then produce bipolar outflows/jets, thus introducing asymmetry to the SN explosion. 
Their study was followed by Ardeljan {\it et al} \cite{Ardeljan_etal-2005} who used a 2D MHD simulation to show that, after an initial stage of linear growth of the toroidal field due to the field-line wrapping by the differential rotation between the proto-NS and the outer envelope, the system enters a second stage marked by an accelerated growth of the magnetic energy due to the~MRI.  They suggested that a growing magnetic pressure could then aid in feeding the bounce shock, essentially transforming it into a fast magnetosonic shock wave that propagates out and disrupts the star in a successful SN explosion. 

Over the last decade, interest in understanding magnetorotational effects on core-collapse SN explosions has grown considerably and this topic now represents a very active area of current research. It has  lead to a large number of new studies, including sophisticated numerical simulations with a realistic equation of state and neutrino transport effects.  In addition to the articles mentioned in the preceding paragraph, other recent studies along similar lines include  \cite{Thompson_etal-2004, Moiseenko_etal-2006, Obergaulinger_etal-2006a, Obergaulinger_etal-2006b, Shibata_etal-2006, Metzger_etal-2007, Uzdensky_MacFadyen-2007a, Uzdensky_MacFadyen-2007b, Burrows_etal-2007b, 
Komissarov_Barkov-2007, Bisnovatyi-Kogan_etal-2008, Mikami_etal-2008, Takiwaki_etal-2009, Obergaulinger_etal-2009, Kuroda_Umeda-2010, Guilet_etal-2011,  Takiwaki_Kotake-2011, Piro_Ott-2011, Masada_etal-2012, Winteler_etal-2012}. 
These MHD magneto-rotational models of SNe  provide a promising avenue towards explaining the bounce shock revival.  However, much more research is needed.  What makes this problem challenging is the fact that, ultimately, these models rely on the production of a strong ($10^{15}$~G or greater) large-scale magnetic field in strongly differentially rotating and turbulent environment. As is well known, a proper computational treatment of MHD turbulence (including that driven by the MRI) and, especially, of large-scale MHD dynamo, requires very high resolution 3D simulations covering a large dynamic range of scales.  
This is challenging even for the most powerful modern super-computers. 
Moreover, the problem is complicated even more by the fact that, as we discussed above, the overall energetics is basically controlled by neutrino heating and cooling processes. Consequently, in order to treat this problem with any degree of confidence, one needs to employ a multidimensional, multi-group neutrino radiation transport scheme that covers both optically thick (inside the NS) and optically-thin regions. This is where the most sophisticated state-of-the-art numerical models are at the present (see, e.g., \cite{Janka-2012, Burrows-2013} for recent reviews).


\subsubsection{Gamma-Ray Bursts (GRBs).}
\label{subsubsec-GRB}

Gamma-ray bursts (see \cite{Piran-1999, Piran-2004, Gehrels_etal-2009} for reviews) are much rarer than SNe and in many ways even more extreme and impressive. 
They shine to us across cosmological distances and are believed to be the most powerful explosions since the Big Bang.  
Observationally, researchers distinguish two separate populations of GRBs --- the short GRBs, with characteristic durations typically less than a couple seconds, and the long GRBs, with durations of 10 seconds or more \cite{Kouveliotou_etal-1993}.  While both classes have comparable luminosities (of order $10^{50}-10^{51} \, {\rm erg/s}$), the long GRBs, because of their longer duration, have larger overall explosion energies. They release  $10^{51}-10^{52}$~ergs of energy in $\gamma$-rays in just a few seconds. 
The short GRBs are believed to come from old stellar populations, and are probably produced in mergers of two compact objects, e.g., two neutron stars or a neutron star and a black hole (see, e.g., \cite{Nakar-2007} for a review). 
The long GRBs, in contrast,  are associated with violent death of some massive stars (perhaps, Wolf--Rayet stars) in younger stellar populations.
The current paradigm, the collapsar model~\cite{Woosley-1993, Paczynski-1998,  
MacFadyen_Woosley-1999, Bromberg_etal-2012}, holds that these  explosions result from the collapse of the core of a massive star, forming a hyper-accreting black hole or a rapidly rotating (millisecond) magnetar acting as the explosion's central engine. 
Thus, the physical environment (densities, temperatures, relevant length- and time-scales) in long GRB central engines is likely to be very similar to that in core-collapse SNe and so most of the processes discussed above in the SN context are probably also relevant to long GRBs. 

The question of the exact nature of the long GRB central engine is very important but is still not settled (see, e.g., \cite{Metzger-2010} for a review).  As mentioned above, there are basically two potentially viable possibilities that have been discussed in the literature.  One is a hyper-accreting (with a mass accretion rate $\dot{M} \sim 1 M_\odot /{\rm sec}$) black hole just formed as a direct result of the core collapse --- this is the so-called failed supernova or "microquasar inside a star" model \cite{Woosley-1993, Paczynski-1998,  MacFadyen_Woosley-1999, Popham_etal-1999, Narayan_etal-2001, DiMatteo_etal-2002, Proga_etal-2003, Matzner-2003, Uzdensky_MacFadyen-2006, Morsony_etal-2007, Barkov_Komissarov-2008, Komissarov_Barkov-2009, Nagataki-2009, Kawanaka_etal-2013}. 
The other model holds that the GRB central engine is a rapidly-rotating (millisecond) magnetar \cite{Usov-1992, Thompson-1994, Blackman_Yi-1998, Zhang_Meszaros-2001, Thompson_etal-2004, Uzdensky_MacFadyen-2007a, Uzdensky_MacFadyen-2007b, Komissarov_Barkov-2007, Metzger_etal-2007, Metzger_etal-2011, Dessart_etal-2008, Bucciantini_etal-2007, Bucciantini_etal-2008, Bucciantini_etal-2009, Takiwaki_etal-2009}.
At the moment, it is not clear how one could resolve this question observationally. 
One promising idea has been recently put forward by Giannios \cite{Giannios-2010}.  He noted that if the central engine is indeed a magnetar, then its magnetic field should be on the high end of the observed magnetar magnetic fields and then one should expect very intense, powerful magnetar activity over the subsequent 100 years or so following the burst. This activity may manifest itself by unusually strong giant SGR flares, so strong that they could be interpreted observationally as short GRBs.  By looking for GRB-explosion remnants in the vicinity of short GRB candidates, this idea can be exploited in practice to settle this question in the foreseeable future. 

Whatever the central engine is, it has to produce an ultra-relativistic jet that burrows its way through the outer stellar envelope  and eventually leads to the observed burst of $\gamma$-rays.  In order to achieve the very high inferred Lorentz factors (of order a few hundred), the jet should have a very high energy per baryon at its base, which translates into a very low level of baryon loading,  about $10^{-5}\, M_\odot$.  Currently, there are two main sets of ideas about what could be the dominant energy carrier in the inner part of the jet:%
\footnote{Free neutrons may also play a very important role in the dynamics and energy release in GRB fireballs \cite{Beloborodov-2003a, Beloborodov-2003b,  Beloborodov-2010}. } 
(1) neutrino models: a very high-entropy, hot (several MeV) photo-leptonic plasma produced as a result of neutrino-neutrino annihilation, e.g., above a collapsar accretion disk;  
and 
(2) magnetic models: Poynting-flux-dominated outflow. 
The advantage of the neutrino model is that, just as in the case of SNe, the temperatures and energy densities in long GRB central engines are so high that the overall energy release is undoubtedly dominated by neutrinos.  However, also as in the case of SNe, the problem with neutrinos is that they couple to the matter only weakly and thus mostly just escape out of the system. As a result, it is not clear whether they are able to deposit enough of their energy at the right place and at the right time to produce a successful jet.

Just as in the case of magneto-rotational models of SNe, Poynting-flux-dominated models of GRBs are motivated by the fact that magnetic fields, even though energetically subdominant to neutrinos, couple to the plasma well. 
Furthermore, most of the models of how astrophysical jets are launched and collimated in other contexts, e.g., for Active Galactic Nucleus (AGN) and Young Stellar Object (YSO) jets, invoke magnetic fields \cite{Blandford_Payne-1982} and are pretty well understood.  Therefore, in one of the leading incarnations of the collapsar model, the jet is launched and collimated magnetically, 
e.g., \cite{Thompson-1994, Meszaros_Rees-1997, Lee_etal-2000, Vlahakis_Konigl-2001, Vlahakis_Konigl-2003, Lyutikov_Blackman-2001, Drenkhahn_Spruit-2002, Lyutikov_Blandford-2002, Lyutikov_Blandford-2003, vanPutten_Levinson-2003, Lyutikov-2006b, Uzdensky_MacFadyen-2006, Uzdensky_MacFadyen-2007a, Uzdensky_MacFadyen-2007b, Burrows_etal-2007b, Komissarov_Barkov-2007, Barkov_Komissarov-2008, Komissarov_Barkov-2009, Bucciantini_etal-2007, Bucciantini_etal-2008, Bucciantini_etal-2009, Dessart_etal-2008, Takiwaki_etal-2009, Nagataki-2009, Komissarov_etal-2009, Lyutikov-2011, Tchekhovskoy_etal-2008, Tchekhovskoy_etal-2010}. 

On general energetic grounds, to be successful, a magnetically-powered GRB jet must have a large-scale magnetic field of order $10^{15}$~G or greater near the central engine  (e.g., \cite{Thompson-1994}), and hence such models are in the regime of interest to this review.  Although very high in absolute terms, the energy density and the pressure of such fields are still small compared with those of the plasma in a newly formed proto-NS or in a collapsar BH accretion disk;  the equipartition level for the magnetic field would be as high as $10^{17}$~G.  Thus, producing $10^{15}$~G fields required for powering GRB jets is energetically quite plausible. 

However, importantly, in order to make a well-collimated jet, what is needed is not just a strong magnetic field but a large-scale field (e.g., with a dipolar topology).  
A  large-scale field of $10^{15}$~G is significantly higher than what one would expect from just the adiabatic compression of a preexisting magnetic flux during the collapse phase.%
\footnote{Except for the case when the pre-collapse progenitor iron core has a magnetic field of order $10^9$~G on a $10^4$~km scale, similar to the most magnetized observed WDs. In this case, however, such a strong magnetic field would brake the collapsing core's rotation during the collapse phase, thus preventing the resulting proto-neutron star from having rapid rotation rate, which is the other required ingredient for GRB.}
Thus, such a strong field has to be produced within the central engine itself, via a large-scale  $\alpha-\Omega$ dynamo, perhaps similar to the way the solar dipolar magnetic field is generated.  MHD dynamo in a rotating turbulent stratified fluid is one of the most important outstanding problems of modern plasma astrophysics. It is the key to understanding the origin of magnetic fields in many different types of objects and environments throughout the universe, including stars, planets, accretion disks, galaxies, etc. It is still not fully understood even in the best-studied case --- our own Sun, although important progress has recently been made (e.g., \cite{Brown_etal-2010}). 

In any case, since it is impossible to produce a net vertical magnetic flux --- i.e., since all the field lines coming out of the central engine must eventually come back to it, --- one expects the large-scale field to have at least dipolar (as opposed to unipolar) topology. The twisting of such a field by differential rotation leads to the build-up of the toroidal magnetic field pressure,  leading to the inflation of the magnetosphere.  Importantly, in the presence of a confining external pressure the inflation is not isotropic but leads to the production of so-called magnetic towers, first proposed by Lynden-Bell \cite{Lynden-Bell-1996, Lynden-Bell-2003} in the context of AGN jets.  The application of the magnetic tower model to core-collapse GRBs was proposed by Uzdensky and MacFadyen for the BH collapsar (\cite{Uzdensky_MacFadyen-2006}) and the millisecond magnetar (\cite{Uzdensky_MacFadyen-2007a, Uzdensky_MacFadyen-2007b}) cases.  The role of the confining external pressure in the GRB case is played effectively by the  inertia of the surrounding stellar material: the rapidly rising magnetic tower launches a shock through the stellar envelope and the pressure of the hot shocked gas in the resulting cocoon provides the lateral confinement for the tower. 
Recent MHD simulations seem to support this picture~\cite{Burrows_etal-2007b, Komissarov_Barkov-2007}. 

Let us summarize  the key plasma-physical problems  important for understanding core-collapse SN  and GRB explosions.  Most of them deal with assessing the plausibility of magneto-rotational explosion mechanisms.  Because the plasmas in these environments are very dense and highly collisional on macroscopic scales, MHD should work well there, and so most of these questions are largely MHD questions, not directly involving kinetic plasma physics. They are: MRI and its large-scale dynamo; magnetic field amplification by differential rotation; resulting field inflation and formation of jets with magnetic tower morphology; stability of the large-scale jet structure. However, some of these MHD processes are tightly linked to magnetic reconnection since they naturally lead to the formation of thin current sheets.  Since reconnection in these current sheets may involve processes that occur on microscopic plasma scales \cite{Uzdensky-2011}, ultimately, understanding microscopic plasma physics beyond MHD is still important in these environments, as we shall discuss again in section~\ref{sec-astro-applications}.  
Furthermore, because magnetic fields involved in these extreme environments are super-critical, 
the plasma physics necessary to describe their behaviour has to be a QED-based plasma physics, which represents the main subject of the next section.


\section{Relativistic quantum plasma physics in ultra-strong magnetic fields}
\label{sec-Quantum-Plasma-Physics} 

Astrophysical problems discussed in the previous section involve plasma environments characterized by extremely high energy densities and, in particular, extremely strong magnetic fields --- fields that approach or exceed the quantum electrodynamic (QED) critical field strength $B_*=m_{\rm e}^{2} c^{3}/e\hbar\cong4.4\times10^{13}\, {\rm G}$.
This is the field strength at which the quantized cyclotron energy $\hbar\omega_{c}$ equals the electron rest energy $m_{e}c^{2} = 511\, {\rm keV}$.  Such environments call for a special theoretical treatment that may differ significantly from classical plasma physics.  In this way, the above extreme plasma-astrophysical problems provide a strong scientific motivation for the development of relativistic quantum plasma (RQP) physics theory in ultra-strong magnetic fields.  Clearly, such a theory has to be based on quantum electrodynamics (QED) as its foundation.  

While the state of development of QED-based RQP theory is still far from being as mature as that of the classical plasma theory, certain progress in this area has been achieved in the past few decades.  The present section is devoted to reviewing the status of the field. 
The section is organized as follows.  First, in section~\ref{subsec-B_*-physics} we outline the various ways in which the presence of a $\sim B_{*}$ field affects the motion of particles and alters the character of plasma behaviour.  
This discussion demonstrates the necessity of quantum-relativistic treatment for ultra-magnetized plasmas.  
Then, we go over the discussion of key conceptual steps in constructing RQP theory: 
first, for  non-relativistic quantum plasma theory (section~\ref{subsec-nonrel-QP}), and then fully-relativistic quantum plasma theory (section~\ref{subsec-rel-QP}).   


\subsection{Basic physics of supercritical magnetic fields}
\label{subsec-B_*-physics}

Magnetic fields approaching $B_{*}$ in magnitude introduce a number of novel features to a plasma.  In this section we first briefly review the motions of single particles in an ultra-strong magnetic field, and then the basic properties of a thermal-equilibrium photo-leptonic plasma with an energy density corresponding to that of the critical magnetic field.  
This is followed by a discussion of light propagation through ultra-magnetized vacuum, 
which is a non-trivial issue due to the effect of vacuum polarization. 
Finally, we discuss equilibrium statistical mechanics of a quantum plasma in an ultra-strong magnetic field.

The discussion is purposefully succint as the material is reviewed thoroughly elsewhere.  
Single particle motions and solutions of the Dirac equation are discussed in detail by Strange~\cite{Strange-1998}.
The cold and warm plasma dielectric tensor for quantizing magnetic fields and radiative processes are reviewed in depth by Harding and Lai~\cite{Harding_Lai-2006}.


\subsubsection{Single particle motion.}

When describing the motion of an electron in a super-critical magnetic field, the necessity of a quantum relativistic treatment can be seen by considering the Landau quantization of electron motion. 
For $B\gtrsim B_{*}$, the Schr\"odinger equation does not suffice and one must solve the Dirac
equation $[\gamma^{\mu} (\partial_{\mu}+ie A_\mu/\hbar c)+{\rmi m_{\rm e}c}/{\hbar}] \Psi=0$ to find the proper Landau orbital energies.
For a uniform magnetic field, the solution of the Dirac equation  reveals energy eigenstates with energies 
\begin{equation}
\left(\frac{E_{r,n,p_{z}}}{m_{\rm e} c^{2}}\right)^{2}=1+\frac{p_{z}^{2}}{m_{\rm e}^{2}c^{2}}+2b(n+r-1),
\label{eq-Dirac_energy}
\end{equation}
where $r=1,2$ differentiates between particle spin states, $n=0,1,2...$ is the energy quantum number, $p_z$ is the momentum component parallel to the magnetic field, and $b\equiv B/B_{*}$ (see, e.g., {\cite{Johnson_Lippmann-1949}).  Thus, even an electron in the ground state has a kinetic energy on the order of its rest mass for $b\simeq 1$.  Equation (\ref{eq-Dirac_energy}) also implies that, unless the plasma temperature exceeds the electron mass by a factor several times unity, most electrons and positrons are confined to the lowest Landau levels.  The interaction of particle spins with the magnetic field is also seen to be significant. For particles with spin magnetic
moment $\bm{\mu_{m}}$, if $\bm{\mu_{m}}\cdot\mathbf{B}\sim k_B T$, the spins may not be randomized and collective spin interactions may introduce novel plasma features.  

For relativistic electrons in a magnetic field, the characteristic length scale is the size of the ground-state wave function \cite{Harding_Lai-2006}, 
\begin{equation}
\rho_{L}=\sqrt{\frac{\hbar c}{eB}}=6.44\times10^{-4}\, (B/{1\, \rm G})^{-1/2}\textrm{cm}, 
\end{equation}
which represents the quantum equivalent of the Larmor radius.
For $B=B_*$ this becomes equal to the (reduced) electron Compton wavelength,
\begin{equation}
\rho_{L*}\equiv \rho_{L}(B=B_*) = 
l_{C} = \frac{\lambda_C}{2\pi} \equiv \frac{\hbar}{m_{\rm e} c} = 3.87 \times10^{-11}\textrm{cm}\,.
\end{equation}
The fact that $\rho_L$ approaches the fundamental quantum length scale $l_C$ calls for the plasma description to be carried into the quantum domain.


\subsubsection{Basic properties of pair plasmas with extreme energy densities corresponding to $B_*^2/8\pi$. }

Critical quantum field introduces a characteristic energy density $u_{*}=B_{*}^{2}/8\pi=7.8\times10^{25}\textrm{erg cm}^{-3}$. 
This high level of energy density indicates the potential for plasmas embedded in such a field to become relativisitcally hot and dense if a substantial portion of the magnetic energy is converted into thermal energy.
For example, a thermal gas of radiation and pairs in thermal equilibrium at zero chemical potential with an energy density $u_{*}$ has a temperature of $k_B T=1.33\,m_{\rm e}c^{2}$, electron and photon number densities of $n_{\rm e}=1.34 \times10^{31}\textrm{cm}^{-3}$ and $n_{\gamma}=9.9\times10^{30}\textrm{cm}^{-3}$
as found by using
\begin{eqnarray}
n_{\rm pair}\left(\theta\right) &=&  
2\int_{mc^{2}}^{\infty}\rmd \epsilon\, a\left(\epsilon\right)\frac{1}{\rme^{\epsilon/k_B T}+1} \nonumber \\
& =& 16\pi\lambda_{\rm C}^{-3}\theta^{3}\int_{\theta^{-1}}^{\infty} \rmd x\sqrt{x^{2}-\theta^{-2}}\, \frac{x}{\rme^{x}+1} \equiv 
16\pi\lambda_{\rm C}^{-3}\theta^{3}I_{1}\left(\theta\right) \, , 
\end{eqnarray}
and
\begin{eqnarray}
u_{\rm pair}\left(\theta\right) &=& 
2 \int_{mc^{2}}^{\infty} \rmd \epsilon\, a_{e}\left(\epsilon\right)\frac{\epsilon}{\rme^{\epsilon/k_B T}+1} \nonumber \\
&=& 16\pi\frac{mc^{2}}{\lambda_{\rm C}^{3}}\theta^{4}\int_{\theta^{-1}}^{\infty} \rmd x \sqrt{x^{2} -\theta^{-2}}\, \frac{x^{2}}{\rme^{x}+1}=
16\pi mc^{2}\lambda_{\rm C}^{-3}\theta^{4}I_{2}\left(\theta\right)\, , 
\end{eqnarray}
\begin{equation}
u_{\gamma}\left(\theta\right)=\frac{8\pi^{5}}{15}mc^{2}\lambda_{\rm C}^{-3}\theta^{4}\, , 
\end{equation}
\begin{equation}
n_{\gamma}\left(\theta\right)=16\pi\zeta\left(3\right)\lambda_{\rm C}^{-3}\theta^{3}\, , 
\end{equation}
where
\begin{equation}
a_{e}\left(\epsilon\right)=8\pi\lambda_{\rm C}^{-3}\frac{\epsilon\, \sqrt{\epsilon^{2}-\left(mc^{2}\right)^{2}}}{\left(mc^{2}\right)^{3}}
\end{equation}
the energy density of states for massive spin $1/2$ fermions \cite{Pathria_Beale-2011}, $\theta\equiv k_B T/m_{\rm e}c^{2}$ and $x\equiv \epsilon/k_B T $, $\rmd \epsilon=k_B T\rmd x$. 

Note that the characteristic inter-particle spacing is then comparable to the reduced Compton wave length:
$l_{\rm int} = n_{\rm e}^{-1/3} \simeq 4.2\times10^{-11}\textrm{cm} = 1.1\, l_C$. 
Since the Compton wave length $\lambda_C = 2\pi l_C$ is the de Broglie wavelength of a particle with kinetic energy equal to its rest energy, we see that the inter-particle spacing in the above-described plasma is comparable to the 
average de Broglie wavelength $\langle\lambda_{\rm dB}\rangle=\langle 2\pi\hbar/p\rangle$.  
This fact, once again, underscores the importance of quantum effects in such a plasma. 

Finally, the inverse plasma parameter, written as the ratio of Coulomb repulsion energy to the average kinetic energy, 
is
\begin{equation}
\Gamma_{\rm p}=\frac{e^{2}}{l_{\rm int}\theta m_{\rm e} c^{2}}=\frac{r_{\rm e}}{l_{\rm int}}\, \theta^{-1} = 
\frac{\alpha}{2\pi}\, \frac{\lambda_{\rm C}}{\theta n^{-1/3}} =
\alpha\left(\frac{I_{1}\left(\theta\right)}{\pi^{2}}\right)^{1/3}\sim\alpha,
\end{equation}
where $\alpha \equiv e^2/\hbar c \simeq 1/137$ is the fine structure constant. 
For example, at $\theta=1.33$, $\Gamma_{\rm p}=0.005$; this justifies the treatment of electrons as a weakly-coupled plasma.  For high temperatures ($\theta \gg 1$) the integral $I_{1}\left(\theta\right)$ asymptotically approaches $3\zeta\left(3\right)/2=1.80$.  As expected, in the ultrarelativistic limit the plasma is weakly coupled due to the
smallness of the fine structure constant~$\alpha$.


\subsubsection{Vacuum polarization and electromagnetic wave propagation in ultra-strong magnetic fields.}
\label{subsubsec-vacuum-polarization}

An ultra-strong field  exerts an important effect on propagation of electromagnetic waves in vacuum, the so-called vacuum polarization (e.g., \cite{Adler-1971, Heyl_Hernquist-1997a, Heyl_Hernquist-1997b, Heyl_Hernquist-2005a, Harding_Lai-2006}).  
In the presence of such fields, virtual particles from the quantum vacuum interact collectively with passing electromagnetic waves, which can be described in terms of one-loop corrections to the effective non-linear QED Lagrangian first derived by Heisenberg and Euler~\cite{Heisenberg_Euler-1936}.  
Vacuum polarization leads to non-linear modifications to Maxwell's equations that can be incorporated as an effective vacuum dielectric tensor~\cite{Heyl_Hernquist-1997b}. 
This leads to a number of important QED effects affecting photon propagation through ultra-magnetized vacuum, such as 1-photon pair creation and  photon splitting (see \cite{Harding_Lai-2006}  for review). 
In addition,  radiative transfer through plasmas in the presence of polarized vacuum is strongly influenced by mode conversion due to a vacuum resonance~\cite{Lai_Ho-2003}.

Heyl and Hernquist \cite{Heyl_Hernquist-1998, Heyl_Hernquist-1999} deduced the effect of an ultrastrong background field on the propagation of Alfv\'en waves using classical MHD coupled with the modified nonlinear Maxwell equations resulting from the effective QED Lagrangian.  They found that the fast magneto-acoustic mode in the nonlinear regime steepens into a shock, inducing pair production at the expense of the wave energy.  They included this process into a model of the phenomena of magnetar flares, where the pair fireball is produced by the breakdown of the QED-modified MHD fast mode~\cite{Heyl_Hernquist-2005a}.  In following works they used the results to explain non-thermal emission produced in more distant regions from the star~\cite{Heyl_Hernquist-2005b,Heyl-2007}.


\subsubsection{Equilibrium quantum statistical mechanics with a uniform magnetic field.}

Electrons obey Fermi-Dirac statistics and thus, in a magnetic field with energy levels given by equation~(\ref{eq-Dirac_energy}), have a distribution function proportional to the occupancy fraction
\begin{equation}
f\left(p_{\Vert},n,r,\mu\right)=\frac{1}{\textrm{exp}\left[\theta^{-1}\sqrt{1+\frac{p_{\Vert}^{2}}{m^{2}c^{2}}+2b(n+r-1)}-\tilde{\mu}\right]+1},\label{eq:magnetized_occupation}
\end{equation}
where $p_{\Vert}$ is the momentum component parallel to the background magnetic field and $\tilde{\mu}=\mu/k_B T$ is the normalized chemical potential. The photons obey Bose-Einstein statistics,
\begin{equation}
f_{\gamma}\left(\omega\right)=\frac{1}{\rme^{\hbar\omega/k_B T}-1}.
\end{equation}

Canuto and Chiu \cite{Canuto_Chiu-1968} obtained the equation of
state for an ideal electron gas in a magnetic field of arbitrary strength
using solutions of the Dirac equation and the energy-momentum tensor
with Lagrangian density 
$\text{\ensuremath{\mathscr{L}}}\equiv-mc^{2}\psi\bar{\psi} - (1/2)\,\hbar c\left(\bar{\psi}\gamma_{\mu}\partial^{\mu}\psi-\partial_{\mu}\bar{\psi}\gamma^{\mu}\psi\right)$, 
where $\psi$ is the (electron) field and $\bar{\psi}$ is the adjoint (positron) field. 
Inserting the spinor solutions of the Dirac equation in a uniform magnetic field and 
assuming a statistical ensemble with weight given by (\ref{eq:magnetized_occupation}), 
they obtained expressions for the perpendicular pressure
\begin{equation}
P_{\bot} = 
2\pi b^{2}\frac{mc^{2}}{\lambda_{\rm C}^{3}} \sum_{n=0}^{\infty} \sum_{r=1}^{2} \int_{-\infty}^{\infty}\frac{\rmd p_{\Vert}}{mc}\, f(p_{\Vert},n,r,\mu)\, \frac{mc^{2}\left(n+r-1\right)}{E\left(p_{\Vert},n,r\right)} \, , 
\end{equation}
the parallel pressure
\begin{equation}
P_{\Vert} = 
2\pi b\frac{mc^{2}}{\lambda_{\rm C}^{3}}\sum_{n=0}^{\infty}\sum_{r=1}^{2}\int_{-\infty}^{\infty}\frac{\rmd p_{\Vert}}{mc}\, f(p_{\Vert},n,r,\mu)\, \frac{p_{\Vert}^{2}/m}{E\left(p_{\Vert},n,r\right)} \, , 
\end{equation}
the energy density
\begin{equation}
U = 
2\pi b\frac{mc^{2}}{\lambda_{\rm C}^{3}}\sum_{n=0}^{\infty}\sum_{r=1}^{2}\int_{-\infty}^{\infty}\frac{\rmd p_{\Vert}}{mc}\, f(p_{\Vert},n,r,\mu)\, \frac{E\left(p_{\Vert},n,r\right)}{mc^{2}} \, , 
\end{equation}
and the number density 
\begin{equation}
n = 
2\pi b\, \frac{mc^{2}}{\lambda_{\rm C}^{3}}\sum_{n=0}^{\infty}\sum_{r=1}^{2}\int_{-\infty}^{\infty}\frac{\rmd p_{\Vert}}{mc}\, f(p_{\Vert},n,r,\mu)\, .
\end{equation}
These expressions can be evaluated numerically for any value of $\theta$ and~$\tilde{\mu}$.

The physical quantities of importance in determining the degree of degeneracy are the density $n$ (or the Fermi temperature $T_{\rm F}$), the temperature $T$, and the chemical potential~$\mu$.  For reference, they are related to each other by the normalization condition (see \cite{Melrose_Mushtaq-2010})  
\begin{equation}
-\, \textrm{Li}_{\frac{3}{2}}\left(-\rme^{\mu/k_B T}\right) =
\frac{4}{3\sqrt{\pi}}\left(\frac{T_{\rm F}}{T}\right)^{3/2}=
\frac{4\pi^{3/2}\hbar^{3}n}{\left(2m k_B T\right)^{3/2}}=
\sqrt{2}\pi^{3/2}n\lambda_{\rm dB}^{3} \, , 
\end{equation}
where $\textrm{Li}$ is the polylogarithm function defined as 
\begin{equation}
\textrm{Li}_{\nu}\left(z\right) \equiv \sum_{k=1}^{\infty}\frac{z^{k}}{k^{\nu}} \, .
\end{equation}


\subsection{Non-relativistic quantum plasma physics}
\label{subsec-nonrel-QP}

As we saw from the discussion in the preceding section, when dealing with plasmas 
immersed in ultra-strong magnetic fields, one has to confront the question of  how to do plasma 
physics while accounting for Landau quantization, quantum statistics, the finite width of the
wavefunction, quantum tunneling and particles with spin. 
This leads into the formalism of \emph{quantum plasma physics.}
This section is intended as a brief overview of the key ideas of non-relativistic quantum plasma physics. 
It also aims at serving as a handy reference to various fundamental results in the theory of quantum plasmas. 
The generalization to relativistic quantum plasmas, applicable to extreme plasma astrophysics, is discussed in section~\ref{subsec-rel-QP}.

The study of collective effects in many-body quantum systems has its roots in condensed matter physics, and much early knowledge in non-relativistic quantum plasma physics comes from that discipline.  Recently, however, quantum plasmas have become popular theoretical and experimental subjects in their own right. 

There has been a significant amount of work on nonrelativistic quantum plasma physics.  
The general procedure employed in modern quantum plasma theory can be outlined as follows. 
The equations of quantum plasma physics typically take the same form as their classical counterparts, but have additional $\hbar$-containting terms.
The basis of modern quantum plasma theory is the Wigner quasi-probability distribution function (WF), through which a kinetic theory is derived.  Defined through the density matrix, the WF is a quantum analog of phase-space density, which facilitates comparison with the classical kinetic theory.
Differentiating the Wigner function with respect to time and using the Schr\"odinger equation, one can obtain the so-called Moyal kinetic equation~\cite{Moyal-1949} that governs the evolution of quantum plasmas.  For the case of particles interacting with a self-consistent mean electromagnetic field, the Moyal equation can be interpreted as a quantum analog of the Vlasov equation. The evolution of the  self-consistent electromagnetic field is then governed by the Maxwell equations as in the classical case.  Having built this general theoretical framework, one can then apply it to study various plasma processes. This, however, has been done only for a few problems, mostly linear waves. 
Next, just as in the classical plasma physics, one can transition from kinetic to fluid description 
by taking the moments of the WF and the Moyal kinetic equation (or its appropriate analog).  
The resulting quantum fluid equations are very similar to the classical fluid equations, except for quantum corrections to the pressure tensor. 

This subsection is devoted to a brief exposition of the nonrelativistic quantum plasma theory. 
In section~\ref{subsubsec-Wigner-Moyal} we present the basic Wigner-Moyal kinetic formalism. 
We illustrate the application of this formalism for linear waves (both electrostatic and electromagnetic) in unmagnetized quantum plasmas in section~\ref{subsubsec-quantum-waves}. 
In section~\ref{subsubsec-nonrel-quantum-fluid}  we discuss the development of the quantum fluid theory and in section~\ref{subsubsec-nonrel-spin-plasma} we discuss spin effects in quantum plasmas.  A very brief review of recent work on nonlinear quantum plasma processes is included in section~\ref{subsubsec-nonrel-nonlinear}.


\subsubsection{The Wigner-Moyal formalism.}
\label{subsubsec-Wigner-Moyal}

In order to build a kinetic quantum plasma theory, one starts by defining the Wigner distribution function~\cite{Wigner-1932}.  The $N-$body Wigner function $f_N$ can be defined in terms of the density matrix $\hat{\rho}$ in the coordinate representation,
$\rho\left(\mathbf{x},\mathbf{x'};t\right) \equiv \langle\mathbf{x}|\psi\left(t\right)\rangle\langle\psi\left(t\right)|\mathbf{x}'\rangle$, as 
\begin{equation}
f_{N}\left(\mathbf{x}_{N},\mathbf{P}_{N};t\right)=\frac{1}{\left(2\pi\hbar\right)^{3N}}\int \rmd^{3N}\mathbf{y}_{N}\,\rho\left(\mathbf{x}_{N}-\frac{\mathbf{y}_{N}}{2},\mathbf{x}_{N}+\frac{\mathbf{y}_{N}}{2};t\right) \textrm{e}^{-\rmi \mathbf{P_{N}\cdot y_{N}/\hbar}}, 
\label{eq:wigfunc1}
\end{equation}
where $\mathbf{x}_{N}$ and $\mathbf{P}_{N}$ are $3N$-dimensional vectors denoting the conjugate positions and canonical momenta coordinates of $N$ particles.  The momentum $\mathbf{P}$ used in this definition is the canonical momentum  $\mathbf{P} = \mathbf{p} + q\mathbf{A}/c$, where $\mathbf{p}= m\mathbf{v}$ is the kinetic momentum of the particle, $\mathbf{v}$ is its velocity, and $\mathbf{A}$ is the electromagnetic vector potential. 
The expectation value of any observable,  $\langle O\rangle = \Tr\left(\hat{\rho}\hat{O}\right)$, 
is then written in the language of the Wigner function as 
\beq
\langle O\rangle = \int \rmd^{3N} \mathbf{x}_{N}\, \rmd^{3N}\mathbf{P}_{N} \, f_{N}\left(\mathbf{x}_{N},\mathbf{P}_{N};t\right) O\left(\mathbf{x}_{N},\mathbf{P}_{N};t\right) \, .
\eeq

Of particular interest in kinetic plasma physics is the 1-particle Wigner function $f\left(\mathbf{x_1},\mathbf{P_1}\right)$, 
\beq
f({\bf x_1,P_1}) =  \int f_{N}\left(\mathbf{x}_{N},\mathbf{P}_{N};t\right)\, \rmd \mathbf{x}_2 ... \rmd \mathbf{x}_N\, 
\rmd \mathbf{P}_2 ... \rmd \mathbf{P}_N = 
{1\over{(2\pi\hbar)^3}}\, \int\limits_{-\infty}^{+\infty}\ \psi^*({\bf x}_1+{\bf y}/2)\, \psi({\bf x}_1-{\bf y}/2)\, 
\rme^{\rmi {\bf P_1\cdot y}/\hbar}\, \rmd^3{\bf y}\, .
\vspace{-5 pt}
\eeq
In the following we will consider only the 1-particle WF and will omit the subscript "1" for brevity. 
In the spirit of Moyal's theory, the evolution of $f\left(\mathbf{x},\mathbf{P}\right)$ is determined from the Schr\"odinger equation, $\rmi\hbar \partial_t \psi = \hat{H}\, \psi $, or the corresponding von Neumann equation for the density matrix, $ \rmi \hbar\, {\partial\hat{\rho}}/{\partial t}=\left[\hat{H},\hat{\rho}\right]$.

The special case of a plasma interacting with a purely electrostatic field, ${\bf A} = 0$ (and hence ${\bf P} = {\bf p}$), allows one to investigate fairly easily important basic kinetic phenomena such as the Landau damping. The Hamiltonian in this case is 
\begin{equation}
\hat{H} = ({\hbar^{2}}/{2m})\nabla^{2} + V\left(\mathbf{x},t\right) = {p^{2}}/(2m)+V\left(\mathbf{x},t\right),
\vspace{-5 pt}
\end{equation}
where $V({\bf x},t) = q \phi({\bf x},t)$, $q$ is the particle charge, and $\phi({\bf x},t)$ is the electrostatic potential.
Then, the evolution equation becomes
\begin{eqnarray}
\frac{\partial f}{\partial t}+\frac{\mathbf{p}}{m}\cdot\nabla f &=& 
\left(2\pi\right)^{-3}\int \rmd^3\mathbf{y} \int \rmd^3\mathbf{s}\,\, \textrm{exp}\left[\rmi\mathbf{y}\cdot\dot{\left(\mathbf{s}-\mathbf{p}\right)}\right]  \nonumber \\
&\times& \rmi q f\left(\mathbf{x},\mathbf{s},t\right)\left[\frac{V\left(\mathbf{r}+{\hbar\mathbf{y}}/{2},t\right) - V\left(\mathbf{r} - {\hbar\mathbf{y}}/{2},t\right)}{\hbar}\right] ,
\label{eq-kinetic-electrostatic}
\end{eqnarray}
which is equivalent to what is known as the Moyal kinetic equation~\cite{Moyal-1949}, 
\begin{eqnarray}
\frac{\partial f\left(\mathbf{x,p,t},\right)}{\partial t} &=& \{\{ f, H \}\} \nonumber \\
&\equiv& -\, \frac{2}{\hbar}\,\textrm{sin}\, \left[ \frac{\hbar}{2}\left(\frac{\partial^{\left(f\right)}}{\partial\mathbf{x}}\frac{\partial^{\left(H\right)}}{\partial\mathbf{p}}-\frac{\partial^{\left(f\right)}}{\partial\mathbf{p}}\frac{\partial^{\left(H\right)}} {\partial\mathbf{x}}\right) \right]\, f\left(\mathbf{x,p},t\right)H\left(\mathbf{x,p},t\right) \,  . 
\end{eqnarray}
Here, $\{\{f,g\}\} \equiv (\rmi\hbar)^{-1}\, (f * g - g* f) = \{f,g\} +O(\hbar^2)$ is the Moyal bracket, $f*g$ is called the Moyal product, and $\{f,g\}$ is the usual Poisson bracket. 
The superscripts $(f)$ and $(H)$ indicate which function the derivative operates on. 
The sine with a differential operator argument is defined in terms of its power series. 
To obtain a complete set of equations, equation~(\ref{eq-kinetic-electrostatic}) is supplemented with the Poisson equation 
\beq
\nabla^2 \phi = -\, 4 \pi \rho({\bf x},t) \, ,
\eeq
where $\rho({\bf x},t)$ is the electric charge density.

In a more general case of a plasma of spinless charged particles interacting with a general electromagnetic field according to the Hamiltonian 
\beq
\hat{H} = {1\over{2m}}\, \left[\mathbf{P} - {q \mathbf{A}\left(\mathbf{x},t\right)/c} \right]^{2} + 
q\phi\left(\mathbf{x},t\right) \, , 
\eeq
a kinetic model based on the WF approach was developed recently 
by Tyshetskiy {\it et al}~\cite{Tyshetskiy_etal-2011}.  
By treating the electromagnetic field as a self-consistent classical mean field as in the classical Vlasov theory (similar to the Hartree mean-field approximation in quantum mechanics), they obtained the following collisionless kinetic equation: 
\begin{equation}
\frac{\partial f\left(\mathbf{x,p},t\right)}{\partial t}+\frac{\mathbf{p}}{m}\cdot\nabla f+q\left[\mathbf{E}+\frac{\mathbf{p\times B}}{mc}\right]\cdot\nabla_{\mathbf{p}}f=I_{q}\left(\mathbf{x},\mathbf{p},t\right) \, ,
\label{eq-kinetic-em}
\end{equation}
where the quantum interference integral $I_{q}\left(\mathbf{x},\mathbf{p},t\right)$ represents the effect of  interference of overlapping wave functions and where we have transitions from using the canonical momentum $\mathbf{P}$ to the kinetic momentum $\mathbf{p} = \mathbf{P} - q \mathbf{A}/c$.  Explicit expression for $I_{q}\left(\mathbf{x},\mathbf{p},t\right)$ in terms of $f(\mathbf{x,p},t)$ and $({\bf A}, \phi)$ reads \cite{Tyshetskiy_etal-2011}:
\begin{multline*}
I_{q}\left(\mathbf{x},\mathbf{p},t\right)=\frac{1}{\left(2\pi\right)^{3}}\int \rmd^3 \mathbf{y}\int \rmd^3\mathbf{s}\,\textrm{exp}\left[\rmi \mathbf{y}\cdot\dot{\left(\mathbf{s}-\mathbf{p}\right)}\right] \\
\times \{ \rmi q f\left(\mathbf{x},\mathbf{s},t\right)\left[\mathbf{y}\cdot\nabla\phi\left(\mathbf{x},t\right)-\frac{\phi\left(\mathbf{x}+\frac{\hbar\mathbf{y}}{2},t\right)-\phi\left(\mathbf{x}-\frac{\hbar\mathbf{y}}{2},t\right)}{\hbar}\right]\\
-\frac{\rmi q}{mc}f\left(\mathbf{x},\mathbf{s},t\right)\left[\mathbf{y}\cdot\nabla\left(\mathbf{s}\cdot\mathbf{A}\left(\mathbf{x},t\right)\right)-\frac{\mathbf{s}\cdot\mathbf{A}\left(\mathbf{x}+\frac{\hbar\mathbf{y}}{2},t\right)-\mathbf{s}\cdot\mathbf{A}\left(\mathbf{x}-\frac{\hbar\mathbf{y}}{2},t\right)}{\hbar}\right] \\
-\,\frac{q}{2mc}\biggl[ \left(\nabla f\left(\mathbf{x},\mathbf{s},t\right)+f\left(\mathbf{x},\mathbf{s},t\right)\nabla\right) \\
+ \frac{\rmi q}{c}f\left(\mathbf{x},\mathbf{s},t\right)\left(\nabla\left(\mathbf{y}\cdot\mathbf{A}\left(\mathbf{x},t\right)\right)-\frac{\mathbf{A}\left(\mathbf{x}+\frac{\hbar\mathbf{y}}{2},t\right)-\mathbf{A}\left(\mathbf{x}-\frac{\hbar\mathbf{y}}{2},t\right)}{\hbar}\right) \biggr] \\
\cdot\left[2\mathbf{A}\left(\mathbf{x},t\right)-\mathbf{A}\left(\mathbf{x}+\frac{\hbar\mathbf{y}}{2},t\right)-\mathbf{A}\left(\mathbf{x}-\frac{\hbar\mathbf{y}}{2},t\right)\right] \} \\
\end{multline*}
Equation (\ref{eq-kinetic-em}) is a quantum analog of the classical Vlasov equation. 


The limits of applicability of the 1-particle WF formalism are set by the weak-coupling condition and by the non-relativistic condition.  Weak coupling can be quantified by specifying that the interaction energy density be much smaller than the kinetic energy density:
$\Gamma=u_{\rm int}/\epsilon_{\rm kin} \simeq 2 e^{2}/(l_{\rm int}\, m\langle v^{2}\rangle) \ll1 $, 
which in the case of a fully degenerate plasma becomes $v_{\rm F}/c \gg \alpha \equiv e^2/\hbar c \simeq 1/137$, 
where $v_{\rm F} \equiv (2 E_{\rm F}/m_e)^{1/2} = (\hbar/m_e)\, (3 \pi^2 n)^{1/3}$ is the Fermi velocity. 
The condition that the non-relativistic approximation is valid for a degenerate plasma is simply $v_{\rm F}/c \ll  1$ as usual.


\subsubsection{Kinetic theory of linear waves and instabilities in unmagnetized quantum plasmas.}
\label{subsubsec-quantum-waves} 
 
An important part of classical plasma physics is the study of linear waves and instabilities. 
The general linear dispersion relation can be written in terms of the linear dielectric permittivity tensor~$\epsilon_{ij}$, which describes the response of a medium to small electromagnetic field perturbations, as 
$\textrm{Det}[\epsilon_{ij}\left(\omega,\mathbf{k}\right)]=0$. 
In an isotropic plasma the dielectric tensor can be divided into longitudinal and transverse parts,
\begin{equation}
\epsilon_{ij}\left(\omega,\mathbf{k}\right)=\epsilon^{long}\left(\omega,\mathbf{k}\right)\frac{k_{i}k_{j}}{k^{2}}+\epsilon^{trans}\left(\omega,\mathbf{k}\right)\left(\delta_{ij}-\frac{k_{i}k_{j}}{k^{2}}\right) \, , 
\end{equation}
For a single-species unmagnetized non-relativistic quantum plasma, the Wigner-Moyal formalism leads to [e.g., \cite{Tyshetskiy_etal-2011, Klimontovich-1980}]
\begin{equation}
\epsilon^{long}\left(\omega,\mathbf{k}\right)=1+\frac{4\pi e^{2}}{k^{2}}\int\frac{\rmd^3\mathbf{p}}{\hbar}\frac{f_{0}\left(\mathbf{p} + {\hbar\mathbf{k}}/2 \right) - f_{0}\left(\mathbf{p} - {\hbar\mathbf{k}}/{2}\right)}{\omega-\mathbf{k}\cdot\mathbf{p}/m},
\end{equation}
\begin{equation}
\epsilon^{trans}\left(\omega,\mathbf{k}\right)=1-\frac{\omega_{\rm p}^{2}}{k^{2}}+\frac{2\pi e^{2}}{m^{2}\omega^{2}}\int\frac{\rmd^3\mathbf{p}}{\hbar}\, p_{\perp}^{2}\, \frac{f_{0}\left(\mathbf{p} + {\hbar\mathbf{k}}/{2}\right) - f_{0}\left(\mathbf{p} - {\hbar\mathbf{k}}/{2}\right)}{\omega-\mathbf{k}\cdot\mathbf{p}/m},
\end{equation}
where $f_{0}$ is the equilibrium Wigner distribution function, $p_{\perp}$ is the component of $\mathbf{p}$ perpendicular to~$\mathbf{k}$, and 
\beq
\omega_{\rm p} \equiv \sqrt{{4\pi n e^2}\over{m}}
\eeq
is the classical plasma frequency.

The generalization of this tensor to magnetized quantum plasmas with a background magnetic field $\mathbf{B}_{0}$ has not yet been carried out. 


\noindent{\bf Electrostatic waves and instabilities:} \\
Electrostatic (e.g., Langmuir) waves leading to a fully-developed plasma microturbulence in ultra-magnetized relativistic quantum plasmas have been invoked by Beloborodov and Thompson \cite{Beloborodov_Thompson-2007}  as a mechanism for effective resistive dissipation of the intense electric currents at the interface between a twisted magnetosphere of an active magnetar and its surface (see section~\ref{subsubsec-magnetars-Xray} and section~\ref{subsec-surface}).  The full theory of how this turbulence develops in the presence of an ultra-strong magnetic field has not yet been developed, so, as a first step, it is important to understand the behaviour of these waves and instabilities in an unmagnetized quantum plasma.  This understanding will hopefully pave the way towards a future more comprehensive theory. 

As an example, the dispersion relation for one-dimensional  longitudinal electrostatic waves in a plasma of quantum electrons with a fixed ion background can be written as (e.g., \cite{Klimontovich_Silin-1960})
\begin{eqnarray}
1 &+& \frac{\omega_{p}^{2}}{k^{2}}\int_{-\infty}^{\infty} \rmd v_{z}\frac{f_{0}\left(v_{z}\right)}{\left(v_{z}-\frac{\omega}{k}\right)^{2}-\left(\frac{\hbar k}{2m_{e}}\right)^{2}} \nonumber \\
&+& \rmi \pi\frac{\omega_{\rm p}^{2}}{k^{2}}\frac{4m}{\hbar k}\left[f_{0}\left(v_{z}+\frac{\hbar k}{2m_{e}}\right)-f_{0}\left(v_{z}-\frac{\hbar k}{2m_{e}}\right)\right]  = 0 \, . 
\end{eqnarray}

For a fully degenerate ($T=0$) Fermi-Dirac plasma, Langmuir (and also ion-acoustic, IAWs) waves were investigated by Eliasson and Shukla~\cite{Eliasson_Shukla-2010}.  They obtained an explicit expansion of the real part of the frequency in powers of $k$ in the long wavelength limit $k\rightarrow 0$, 
\begin{equation}
\omega^{2}=\omega_{\rm p}^{2}+\frac{3}{5}k^{2}v_{\rm F}^{2}+\left(1+\frac{48}{175}\frac{m^{2}v_{\rm F}^{2}}{\hbar^{2}\omega_{\rm p}^{2}}\right)\frac{\hbar^{2}k^{4}}{4m^{2}}.
\end{equation}
and also mentioned the possibility of Landau damping even in the completely degenerate case, although did not present a specific calculation or expression for it. 

Melrose and Mushtaq \cite{Melrose_Mushtaq-2010}  applied for the first time the longitudinal dielectric tensor to electrostatic waves in plasmas of arbitrary degeneracy. They examined the fully degenerate and non-degenerate limits for Langmuir waves and IAW's in more detail, utilizing a power series expansion in the degeneracy parameter $\exp(\mu/k_B T)$, and derived both the real and imaginary parts of the frequency, as well as the screening length.

Recently, Rightley \& Uzdensky (2014, in preparation) developed a general semi-analytical procedure for studying linear electrostatic waves and instabilities in unmagnetized quantum plasmas with arbitrary equilibrium distribution, including arbitrary degree of degeneracy, drifting sub-populations, etc.  This provides a single coherent framework for exploring a wide range of kinetic phenomena, including Buneman, bump-on-tail, ion-acoustic instabilities, over broad parameter ranges, and for investigating the role of quantum effects.


Electrostatic streaming instabilities, such as two-stream and bump-on-tail, have been also investigated for nonrelativistic quantum Schr\"odinger-Poisson plasmas  using a different approach, namely, a multi-stream model cast in terms of Cauchy distributions, which yielded analytically tractable results~\cite{Haas_etal-2000, Haas_etal-2001}. 
It was established that no analog of the Penrose functional exists for electrostatic waves in quantum plasmas, though the stability of single-peaked distribution functions remains valid. 
Haas and Bret \cite{Haas_Bret-2012}  investigated the Buneman instability using a cold fluid model with collisions introduced phenomenologically.  They found that the presence of the quantum Bohm pressure term tended to stabilize the system in the linear regime and led to oscillatory behaviour in the nonlinear regime not present in the purely classical case.
General, realistic studies with Fermi-Dirac electrons of arbitrary degeneracy are yet to appear. 


\noindent{\bf Electromagnetic Waves and Instabilities:} \\
The dispersion relation for electromagnetic waves propagating through a non-relativistic quantum plasma of fully degenerate electrons and stationary ions, 
\begin{equation}
\omega^{2}=\omega_{\rm p}^{2}+\left(kc\right)^{2} + ({\hbar k^2}/{2m})^{2},
\end{equation}
was obtained by Zhu and Ji \cite{Zhu_Ji-2012} as a non-relativistic limit of a more general expression. 
Dispersion relations for electromagnetic waves in  plasmas of arbitrary degeneracy have yet to appear in the literature.

Bret and Haas \cite{Bret_Haas-2011} 
derived the full electromagnetic response tensor for a non-relativistic quantum plasma and applied it to  the onset and growth of the electromagnetic filamentation instability. 
Linear Weibel instability was studied by Haas \cite{Haas-2008}, who found that quantum effects tend to weaken or suppress the instability, and also by Tsintsadze \cite{Tsintsadze-2009}, who found a novel oscillatory Weibel instability. Haas {\it et al} \cite{Haas_etal-2009} also studied the nonlinear saturation of the quantum Weibel instability numerically.


\subsubsection{Non-relativistic quantum fluid theory.}
\label{subsubsec-nonrel-quantum-fluid}

As in the classical case, the kinetic description of quantum plasmas often carries more information than is actually needed to adequately describe certain phenomena.  The relative simplicity of fluid theory is often worth the sacrifice in the domain of validity, especially when one deals with nonlinear physics.  This justifies developing a fluid theory, using fluid quantities defined via the moments of the WF 
\begin{equation}
n\left(\mathbf{x},t\right)=\int \rmd \mathbf{P}\, f\left(\mathbf{x},\mathbf{P};t\right) \, , 
\label{eq-quantum-fluid-n}
\end{equation}
\begin{equation}
\mathbf{u}\left(\mathbf{x},t\right)=\frac{1}{mn} \int \rmd\mathbf{P} \, f\left(\mathbf{x},\mathbf{P};t\right)
\left(\mathbf{P} - q \mathbf{A}/c\right) \, , 
\label{eq-quantum-fluid-u}
\end{equation}
\begin{equation}
\mathbf{\Pi} \left(\mathbf{x},t\right) = \frac{1}{m^{2}}\int \rmd \mathbf{P}\, f\left(\mathbf{x},\mathbf{P};t\right)\left(\mathbf{P} 
- q\mathbf{A}/c\right) \otimes \left(\mathbf{P}-q\mathbf{A}/c\right) - nm\, \mathbf{u}\otimes\mathbf{u} \, , 
\label{eq-quantum-fluid-pressure}
\end{equation}
that obey evolution equations systematically derived by taking the moments of the quantum kinetic (e.g., Moyal) equation.
Thus, Manfredi and Haas \cite{Manfredi_Haas-2001} derived a set of quantum hydrodynamic (QHD) equations and then Haas \cite{Haas-2005}  obtained quantum magnetohydrodynamic (QMHD) equations.  It was found that the continuity equation,  
\beq
\partial_{t} n + \nabla\cdot\left(n\mathbf{u}\right) = 0 \, , 
\label{eq-quantum-fluid-continuity}
\eeq 
is the same as in the classical case. 
The momentum equation, assuming for simplicity a scalar pressure, $\Pi_{ij} = P\delta_{ij}$, becomes 
\begin{equation}
\partial_{t}\mathbf{u}+\left(\mathbf{u}\cdot\nabla\right)\mathbf{u} = 
\frac{q}{m}\left(\mathbf{E}+{{\mathbf{u}\times\mathbf{B}}\over{c}}\right)-\frac{1}{mn}\nabla P+\frac{\hbar^{2}}{2m^{2}}\nabla\left(\frac{\nabla^{2}\sqrt{n}}{\sqrt{n}}\right) 
\label{eq-quantum-fluid-momentum}
\end{equation}
and is identical to its classical counterpart except for the additional pressure term proportional to~$\hbar^{2}$.
This quantum term can be important when there are structures on the scale of the thermal de Broglie wavelength  and when the parameter $\hbar\omega_{\rm p}/k_{\rm B} T$ approaches or exceeds unity. 
Haas \cite{Haas-2005} goes on to discuss two-fluid QMHD with phenomenological collisional terms, ideal QMHD, and a class of magnetostatic QMHD equilibria.

Fluid theory for quantum plasmas is still an actively-developing and controversial topic; a criticism of quantum plasma fluid theory was recently provided by Vranjes {\it et al} \cite{Vranjes_etal-2012}.


\subsubsection{Non-relativistic quantum spin-1/2 plasma (Pauli plasma).}
\label{subsubsec-nonrel-spin-plasma}

Particle spin introduces a new level of complexity into the dynamics of plasmas, especially in magnetized plasmas. 
In non-relativistic quantum mechanics for spin-1/2 particles, spin is incorporated 
by generalizing the wave-function to a two-component spinor $\left(\Psi_{\uparrow},\Psi_{\downarrow}\right)$
corresponding to ``up'' and ``down'' spin components, respectively.
For electrons (with charge $q=-e$), the evolution equation for this spinor in a magnetic field $\mathbf{B}$ is the Pauli equation: 
\begin{eqnarray}
\rmi \hbar\, \partial_{t}\Psi &=& \hat{H}\, \Psi \nonumber \\
&=& \biggl[ - \frac{\hbar^{2}}{2m}\left(\nabla+\frac{\rmi e}{\hbar c}\mathbf{A}\right)^{2}\mathbf{I} 
- e \phi + \mu_{\rm B}\mathbf{B}\cdot\bm{\sigma} \biggr]\, \Psi
\label{eq-Pauli-1}
\end{eqnarray}
where 
$\mathbf{I}$ is the $2\times2$ unit matrix, 
$\bm{\sigma}$ is the vector of Pauli matrices,
$\bm{\sigma}=\left(\sigma_{x},\sigma_{y},\sigma_{z}\right)$, 
and $\mu_{\rm B}$ is the electron magnetic moment $e\hbar/2m_{\rm e}$. 
It should be noted that this equation neglects direct spin-spin interactions,
which can influence wave dispersion and may generate macroscopic-scale magnetic fields.
These interactions can in principle be included with the more general Hamiltonian
\begin{equation}
\hat{H}= -\, \frac{\hbar^2}{2m}\left[\bm{\sigma\cdot}\left(\nabla+\frac{\rmi e}{\hbar c}\mathbf{A}\right)\right]^{2}-e\phi.
\end{equation}
In an ultrastrong magnetic field, however, spin-spin interactions can be expected to be dominated by the external field.

Marklund and Brodin \cite{Marklund_Brodin-2007} used the Pauli equation along with a Madelung decomposition of the wavefunction to derive a set of quantum spin magnetized fluid equations, referred to as the Spin Quantum Hydrodynamics (SQHD).
In addition to the usual fluid quantities (the density, velocity, and the pressure tensor), the quantum spin fluid theory includes one more vector,  the spin density $\mathbf{S}_{s}$ for species $s$, and hence one more equation --- the one governing the evolution of~$\mathbf{S}_{s}$:
\begin{equation}
\left(\partial_{t}+\mathbf{u}_{s}\cdot\nabla\right)\mathbf{S}_{s}=
-\, \frac{2\mu_{s}}{\hbar}\, \mathbf{B}\times\mathbf{S}_{s} \, .
\end{equation}
In addition, the interaction of the spin density vector with the magnetic field results in an extra force in the momentum equation, $(2n_s \mu_s /\hbar)\, S_s^j \nabla B_j$, where $n_s$ and $\mu_s$ are the number density and the magnetic moment of the particles of species~$s$.  
Otherwise, the set of fluid equations is identical to the one for spinless quantum plasmas [see equations (\ref{eq-quantum-fluid-continuity})-(\ref{eq-quantum-fluid-momentum}) above], including the quantum pressure-like term $\hbar^2 /(2m^2)\, \nabla (n^{-1/2}\,{\nabla^{2}\sqrt{n}})$ in the equation of motion. 

Marklund and Brodin \cite{Marklund_Brodin-2007}  then applied their spin fluid theory to derive the linear susceptibility tensor
in the low frequency ($\omega/\Omega_{\rm c}\ll1$, where $\Omega_{\rm c}$ is the cyclotron frequency) and highly magnetized ($\mu B_{0}/k_B T\gg1$) limits. 
They found that spin can significantly alter the dispersion of the fast and slow magnetosonic waves, which  may have important implications for cold laboratory plasmas and also for ultra-magnetized astrophysical plasmas.  
This spin fluid model was subsequently also applied to study the influence of spin on Alfv\'en waves~\cite{Brodin_etal-2008}.  It was found that, for wave phenomena occurring on time scales shorter than the spin-flip timescale but longer than the electron cyclotron time, spin-up and spin-down electrons essentially behave as two separate fluids, which influences the nonlinear aspects of the wave propagation.


A recent critique of some of the SQHD theories of spin-1/2 quantum plasmas was offered by Krishnaswami {\it et al}~\cite{Krishnaswami_etal-2013}.


\subsubsection{Nonlinear processes in quantum plasmas.} 
\label{subsubsec-nonrel-nonlinear}

Nonlinear processes in quantum plasmas have been the focus of a great deal of interest in recent years, motivated by current and near-future experiments with high-intensity lasers \cite{Marklund_Shukla-2006, Mourou_etal-2006} and by the potential for application to dense astrophysical plasmas.
The study of nonlinear quantum plasma phenomena has so far concentrated mostly on non-linear wave propagation (see, e.g., \cite{Shukla_Eliasson-2010} for a  review of nonlinear  solitary waves in electrostatic quantum plasmas). 
Akhtar and Mahmood \cite{Akhtar_Mahmood-2011}  investigated the formation of weakly nonlinear double layers in non-relativistic electron-positron-ion plasmas using the QMHD equations and obtained  a Kortewig-deVries-like equation that describes this process.
More complex nonlinear processes, such as turbulence or reconnection, have not yet been studied in quantum plasmas, even non-relativistic ones.


\subsection{Relativistic quantum plasma physics}
\label{subsec-rel-QP}

Basic quantum mechanics is not capable of accounting for all the physics occuring in the presence of supercritical fields, and one needs to turn to a relativistic quantum field theory (QFT). 
In QFT one deals with second quantized field operators instead of complex-number wavefunctions. 
Correspondingly, relativistic quantum plasma physics extends the Wigner function formalism to relativistic systems in terms of quantized fields. The WF is then replaced by a Wigner operator. 
For physics excluding nuclear matter and neutrinos, the applicable QFT is quantum electrodynamics (QED). 
For spin-$1/2$ fermions of QED the Wigner operator is a $4\times4$ matrix, built from a direct product of two Dirac bispinors corresponding to spin-up and spin-down particles and antiparticles.

The main applications of relativistic quantum plasma physics include laboratory laser-produced pair plasmas; quark-gluon plasmas, e.g., in relativistic heavy-ion collision experiments (although this area lies beyond QED); and various astrophysical applications described in this review. 

A complete description of a plasma in terms of QED is difficult to incorporate in the already complex problems of plasma physics.  However, some key phenomena predicted by QED have been isolated and included in a number of studies in the context of ultra-magnetized neutron stars.  Such phenomena include polarization of virtual pairs leading to non-trivial optical behaviour of the vacuum (vacuum polarization, see section~\ref{subsubsec-vacuum-polarization}), anisotropy in Compton and Coulomb scattering, and modifications to the emission of synchrotron radiation. 

In this subsection we outline the recent progress in building the fundamental theoretical framework of relativistic quantum plasma physics, first for spinless (Klein-Gordon) particles (section~\ref{subsubsec-KG}) and then for spin-1/2 (Dirac) fermions (section~\ref{subsubsec-KG}).


\subsubsection{Relativistic quantum kinetic theory without spin (Klein-Gordon plasma).}
\label{subsubsec-KG}
 
Mendonca \cite{Mendonca-2011} used a relativistic scalar Wigner function and the Klein-Gordon (KG) equation 
$-\,c^{-2}\, \partial_t^2 \psi + \nabla^2\psi  =  (mc/\hbar)^2\, \psi$,  
to derive a relativistic quantum Vlasov kinetic equation for unmagnetized plasmas composed of spinless (scalar) charged particles (Klein-Gordon plasma), treating electromagnetic field as a classical mean field. 
He then used it to derive the dispersion relations for longitudinal and transverse linear waves. 
In the high frequency limit, he arrived at specific forms for the dispersion relation for longitudinal electrostatic and transverse electromagnetic waves for degenerate electrons in a stationary neutralizing ion background:
\begin{equation}
\omega_{\rm L}^{2}=\frac{\omega_{\rm p}^{2}}{\gamma_{\rm F}} \left[1+\frac{k^{2}}{\omega^{2}}\left(1-\frac{\omega^{2}}{k^{2}c^{2}}\right)v_{\rm F}^{2} + \left(\frac{\hbar k}{2m}\right)^{2}\gamma_{\rm F}^{-2}\right] \, , 
\end{equation}
\begin{equation}
\omega_{T}^{2}=\left(kc\right)^{2}+\frac{\omega_{\rm p}^{2}}{\gamma_{a}^{2}}
\end{equation}
where
\begin{equation}
\gamma_{\rm F}\equiv\langle\gamma\rangle=\int\gamma f_{0}\left(\mathbf{v}\right) \rmd \mathbf{v}=\sqrt{1+v_{\rm F}^2/c^2},
\end{equation}
and  
\begin{equation}
\gamma_{a}\equiv\sqrt{1+\frac{eA_{0}}{mc}} \, ,
\end{equation}
with $A_{0}$ being the amplitude of the (circularly polarized) wave.

In the weakly-relativistic limit $\beta_{\rm F}\equiv v_{\rm F}/c\ll1$, analogous results were also obtained by Zhu and Ji \cite{Zhu_Ji-2012} who started from the general kinetic equation for the Dirac plasma and then neglected the spin terms. They obtained:
\begin{equation}
\omega_{\rm L}^{2}=\omega_{\rm p}^{2}\left(1-\frac{1}{2}\beta_{\rm F}^{2}\right)+\frac{3}{5}\left(kv_{\rm F}\right)^{2}\left(1-\frac{3}{2}\beta_{\rm F}^{2}\right)+\left(\frac{\hbar k}{2m}\right)^{2}\left(1-\frac{3}{2}\beta_{\rm F}^{2}\right),
\end{equation}
\begin{equation}
\omega_{\rm T}^{2}=\left(kc\right)^{2}+\omega_{\rm p}^{2}\left(1-\frac{1}{2}\beta_{\rm F}^{2}\right)+\left(\frac{\hbar k}{2m}\right)^{2}\left(1-\frac{3}{2}\beta_{\rm F}^{2}\right).
\end{equation}
However, the case of a strongly magnetized Klein-Gordon plasma has not yet been properly worked out.  It is still not clear how this formalism can be used for strongly-magnetized case, where the perpendicular motion of particles is quantized in discrete Landau states.


\subsubsection{Relativistic quantum kinetic theory with spin 1/2 (Dirac plasma).}
\label{subsubsec-Dirac}

Relativistic quantum plasma theory for Dirac spin-1/2 fermions has so far dealt mostly with linear waves. 
The physics involved is important, in particular, for the accurate modeling of observable radiation from ultra-magnetized astrophysical plasmas, e.g., in the context of magnetars. 

Melrose \cite{Melrose-2008} has reviewed in detail many fundamental aspects of waves in unmagnetized relativistic plasmas using the full QED formalism.  He has carried out a generalization of vacuum QED  to quantum electrodynamics of the self-consistent electromagnetic field of a plasma. 
His theory expands the response of the medium in a series of response tensors of various orders, allowing a treatment of basic wave propagation in the linear regime and also of nonlinear wave-wave interactions and wave-particle interactions, including the influence of the plasma on QED scattering processes.  
The response tensor formalism was generalized also to the case of a magnetized RQP \cite{Melrose_Weise-2009, Melrose_Weise-2012}. 

However, in order to understand fully nonlinear quantum plasma physics, a method not based on perturbative expansion is desirable. Such a method can be found in relativisitic quantum kinetic theory and in quantum fluid theories.

Hakim and Sivak \cite{Hakim_Sivak-1982} developed a covariant Wigner function approach for the Dirac plasma in a strong magnetic field and derived the relativistic quantum Liouville equation for the corresponding Wigner operator. 
They obtained the equation of state for an ensemble of particles in thermal equilibrium, wrote down the Wigner function for the QED vacuum to make connections with vacuum polarization, and discussed the fluctuations of one-particle quantities and their spectrum.
Going beyond the mean-field approximation, Hakim {\it et al} \cite{Hakim_etal-1992} extended the Wigner function approach for the Dirac plasmas using the relaxation-time approximation. Through the Chapman-Enskog procedure,  they calculated the basic transport coefficients as functions of the temperature, density, spin polarization, and the assumed relaxation times.


\subsubsection{Relativistic hydrodynamics.}
\label{subsubsec-rel-hydro}

While classical relativistic ideal fluid theory is well understood and has been applied to many astrophysical situations, relativistic viscous hydrodynamics and relativistic quantum fluids have been studied only much more recently, mostly in the context of the quark-gluon plasma produced in heavy ion collisions for the case of non-zero viscosity (see~\cite{Romatschke-2010} and for laser-produced quantum plasmas.
More recently, a hydrodynamic model for relativistic Dirac spin-1/2 quantum plasmas was developed~\cite{Asenjo_etal-2011}.


\section{Astrophysical applications} 
\label{sec-astro-applications}

This section discusses how the understanding of plasma physics in extreme environments described in the previous section can be applied to actual astrophysical systems. In particular, we  consider five issues: 
(1) the propagation of photons across an ultra-strong magnetic field in a magnetar magnetosphere; 
(2) large-scale MHD processes in central engines of SNe and GRBs and in magnetar magnetospheres; 
(3) thermodynamics of extreme plasma environments; 
(4) interaction of strong magnetospheric currents in an active twisted magnetar magnetosphere with the NS surface, possibly responsible for the production of hard X-ray emission; 
and 
(5) reconnection in magnetar magnetospheres as a mechanism powering giant $\gamma$-ray flares in Soft Gamma Repeaters (SGRs) and possibly GRBs.


\subsection{Propagation of light through magnetospheres of magnetars}
\label{subsec-light-propagation}

The topic of light propagation through an ultra-magnetized plasma in magnetar magnetospheres is important for two reasons.  First, it affects radiative processes, such as radiative cooling and particle creation, that control the physical conditions in, e.g., a reconnection layer in SGR flares, see section~\ref{subsec-reconnection}.
Second, since direct observation of radiation is often our only diagnostic tool (which is true in astrophysics in general), 
our interpretation of various phenomena in magnetars is subject to modification due to the propagation of the photons to us across the magnetosphere. 

In sharp contrast to the case of weak magnetic fields, an ultra-strong field in a magnetar magnetosphere (and, to a lesser extent, slightly subcritical, $\sim 10^{12}\, {\rm G}$, fields in magnetospheres  of normal NSs) strongly affects the propagation of high-energy photons even in a vacuum, in the absence of plasma. 
This is due to the so-called QED vacuum polarization effect or vacuum birefringence reviewed in section~\ref{subsec-B_*-physics}.  
The effect arises due to the interaction of light with virtual pairs in the presence of the strong magnetic field and corresponds to one-loop corrections in the QED (Euler-Heisenberg) Lagrangian. There has been a lot of work on understanding the effect of vacuum polarization on electromagnetic waves, motivated by both fundamental physics and astrophysics (e.g., magnetar) applications (see, e.g., \cite{Heyl_Hernquist-1997a, Heyl_Hernquist-1997b}). 

Even though, formally speaking, the propagation of photons through a QED vacuum is not within the scope of plasma physics, it is still important to review it because it changes the energy-exchange processes between radiation and pairs. 
Overall, the picture can be summarized as follows [see, e.g., \cite{Harding_Lai-2006} for review]. 
First, it is important to distinguish between the two photon polarizations (for non-parallel propagation): the ordinary mode (the O-mode), whose electric field vector has a component parallel to the background magnetic field, and the extraordinary mode (the X-mode), whose electric field is perpendicular to the background magnetic field. 
The propagation of X-mode photons across the field is strongly suppressed by the QED process of photon splitting. 
The attenuation coefficient for this process becomes insensitive to $B$ for $B> B_*$, but is a strong function of the photon energy ($\sim \epsilon_{\rm ph}^5$).  This means that, basically, only photons with energies $\epsilon_{\rm ph} \lesssim 0.1\, m_{\rm e} c^2 \simeq 50\, {\rm keV}$ are able to escape the strong-field region of size $L\sim R_{\rm NS} \sim 10^6\, {\rm cm}$ around a magnetar.  In addition, X-mode photons with energies exceeding the threshold for pair creation with one particle created at the zeroth and the other at the first excited Landau level [this threshold is $\epsilon_{\rm ph} = (1+\sqrt{1+2b})\,m_{\rm e} c^2 \simeq 3 \, {\rm MeV}$ for $b=10$] are subject to additional attenuation due to 1-photon pair production. 
For these pair-producing photons this process is even more important than photon splitting. 
The process of 1-photon pair production also leads to heavy absorption of pair-producing (MeV) O-mode photons.  This process and its inverse essentially ensure that these MeV photons are in equilibrium with pairs in environments with high plasma energy density, e.g., inside a reconnection layer during an SGR flare. In contrast, the photon splitting process does not affect O-photons.  Sub-MeV O-photons thus can propagate essentially freely across a vacuum magnetic field.  

An ultra-strong magnetic field also affects the nonlinear behaviour of vacuum electromagnetic  waves.  In particular, Heyl and Hernquist \cite{Heyl_Hernquist-1998, Heyl_Hernquist-1999} found that, nonlinearly, electromagnetic waves steepen and tend to form shocks as they propagate across an ultra-strong magnetic field. 

QED effects associated with an ultra-strong magnetic field are also important to take into account when considering electromagnetic radiation propagating through a plasma. In particular,  an ultra-strong field affects various photon-particle interaction processes, such as Thomson/Compton scattering, photon-photon pair creation and annihilation, and cyclotron/synchrotron emission and absorption (e.g., \cite{Herold-1979,  Arons_Barnard-1986, Baring-1988, Lyutikov_Gavriil-2006, Harding_Lai-2006}).
For example, it was shown that Compton scattering of X-photons is strongly suppressed by the magnetic field: $\sigma_{\rm es}^{(X)}/\sigma_{\rm T} \simeq (\epsilon_{\rm ph}/b\, m_{\rm e} c^2)^2$ for a photon with energy $\epsilon_{\rm ph} \ll b\, m_{\rm e} c^2$ propagating perpendicular to the magnetic field~\cite{Herold-1979}.  In contrast,  the Compton scattering of O-mode photons is not strongly suppressed and so this process often provides the main source of opacity for sub-MeV O-mode photons.  The dispersion relation for normal electromagnetic modes propagating in a relativistic pair plasma in an ultra-strong field was derived by Arons and Barnard~\cite{Arons_Barnard-1986} in the context of neutron star magnetospheres.  Resonant cyclotron scattering and Comptonization in a non-relativistic warm plasma immersed in an inhomogeneous magnetic field, as well as their observational implications for AXPs, were analyzed by Lyutikov and Gavriil~\cite{Lyutikov_Gavriil-2006}. 

All these processes are important to take into account when modeling observable radiative signatures of high-energy processes in magnetospheres of magnetars and, to some degree, even of normal~NSs.


\subsection{Large-scale MHD processes and current sheet formation leading to reconnection onset}
\label{subsec-MHD}

As we have seen in section~\ref{sec-extreme-astro-environments}, many of the plasma-physical aspects of magnetar flares and SN/GRB central engines involve large-scale MHD dynamics believed to be similar to MHD processes taking place in various other astrophysical systems across the Universe.  
For example, MHD dynamo in a rotating convective star is likely to be the main mechanism for the generation of magnetic fields both in young NSs and in main-sequence stars, such as the Sun.  
In addition, $\gamma$-ray flares in active SGR systems may happen in a manner dynamically similar to solar flares.  In particular, they may involve magnetospheric field-line twisting by foot-point motions on the NS crust, subsequent inflation of magnetic flux ropes leading to the loss of equilibrium and/or the development of the kink instability, which creates current sheets and thus sets up the macroscopic conditions for reconnection.  
Likewise, MHD processes in long-GRB collapsar central engines probably share a lot of similarities with other BH accretion-disk/jet systems such as XRBs and AGNs. For example, MRI turbulence may provide a mechanism for angular momentum transport in a collapsar accretion disk and, at the same time, through its MHD dynamo action, generate a large-scale magnetic field.  The latter then facilitates the launching, collimation, and acceleration of a relativistic magnetic tower jet. 
In the millisecond magnetar scenario for GRBs, a similar magnetic-tower-like jet may be collimated by the interaction of the rapidly rotating magnetar magnetosphere with the surrounding infalling stellar envelope, as has been illustrated, e.g., with the "pulsar-in-a-cavity" model~\cite{Uzdensky_MacFadyen-2007a, Uzdensky_MacFadyen-2007b}, or with the relativistic pulsar-wind model \cite{Bucciantini_etal-2007, Bucciantini_etal-2008, Bucciantini_etal-2009} (see  section~\ref{subsec-SN_GRB} and also~\cite{Komissarov_Barkov-2007}).

Importantly, in order to be successful, the large-scale magnetic field in a magnetic-tower (or any other magnetically-powered) GRB jet should be able to survive the propagation through the progenitor star. Determining under what conditions this is at all possible and in what form (i.e., with what kind of structure) the magnetic energy emerges at large distances, is an important and nontrivial problem of extreme plasma astrophysics. 
The reason why this is a difficult issue is that there are  large-scale MHD processes that present strong obstacles to the jet propagation.  Generically, the magnetic field in a GRB jet may develop spatial and temporal intermittency and have a complex substructure~\cite{Uzdensky_MacFadyen-2007a, Uzdensky_MacFadyen-2007b, McKinney_Uzdensky-2012}.  The evolution of this substructure may lead to the formation of thin current-sheet configurations that can be classified according to the type of the underlying magnetic geometry, as was discussed by McKinney and Uzdensky~\cite{McKinney_Uzdensky-2012}.
For example, field-line opening in an inflating magnetosphere driven by differential rotation (e.g., \cite{Uzdensky-2002a, Uzdensky-2004}) or by relativistic rotation beyond the light cylinder (e.g., \cite{Spitkovsky-2006, Contopoulos_Kalapotharakos-2010, Tchekhovskoy_etal-2013}) may naturally lead to the formation of thin current sheets, across which magnetic field changes sharply. 
In addition, it is generally quite unlikely that the large-scale magnetic field driving the jet at the base would have a simple, e.g., dipole-like, geometry maintained for the duration of the event.  For example, if the field is produced by the MRI dynamo in a collapsar accretion disk, then, as numerical simulations of MRI turbulence show~\cite{Davis_etal-2010}, the toroidal field emerging above the disk into the corona changes its polarity in a quasi-periodic fashion on a time scale of about 10 orbits. This implies that the toroidal field in the jet pumped over the entire (several seconds) duration of the event may experience hundreds of direction reversals and, correspondingly, may contain hundreds of magnetic "stripes" separated by current sheets (similar to the striped pulsar wind~\cite{McKinney_Uzdensky-2012}). 
Finally, even if the magnetic field  near the central engine is somehow well-ordered, it may develop irregularities over time as it propagates out.  In particular, the highly-twisted magnetic field in the jet can develop ideal MHD instabilities, such as the kink~\cite{Drenkhahn_Spruit-2002, Giannios_Spruit-2006, Uzdensky_MacFadyen-2007a}, leading to current sheet formation and reconnection onset, similar to sawtooth crashes in tokamaks. 

Importantly, current sheets thus produced become potential sites for the onset of magnetic reconnection, which therefore  becomes an important issue for magnetically-driven GRB models.
The effectiveness of reconnection governs where and whether the Poynting flux in the jet is dissipated and thus determines the fate of the jet, i.e., whether it crumbles or survives, with important observational consequences~\cite{Lyutikov_Blackman-2001, Lyutikov_Blandford-2002, Lyutikov-2006b, Giannios_Spruit-2006, Giannios_Spruit-2007, McKinney_Uzdensky-2012}. 
In particular, if reconnection proceeds successfully early on, e.g., while the jet is still making its way through the progenitor star, then it may lead to the break-up of a single large-scale magnetic structure (like a magnetic tower) into a train of multiple smaller toroidal plasmoids~\cite{Uzdensky_MacFadyen-2007a}.  This picture is rather similar to the cyclic behaviour involving reconnection suggested for magnetospheres of accreting young stars~\cite{Goodson_etal-1997, Uzdensky-2004}. The typical sizes of such  plasmoids and their production rate may then influence the variability of dissipation at later times and larger distances in the GRB jet, leading to observable timing signatures.  In turn, they are controlled by detailed reconnection dynamics and determining them requires a thorough understanding of reconnection process in this extreme environment, which we will  discuss  in section~\ref{subsec-reconnection}.


\subsection{Thermodynamics of extreme plasma environments}
\label{subsec-thermo}

Despite this general similarity in the macro-scale MHD {\it dynamics} in active magnetar magnetospheres and central engines of SNe and GRBs on the one hand, and the more familiar and well-studied solar corona and accretion disk/jet systems on the other, the {\it thermodynamics} of these two types of environments is very  different. Many of the differences are due to the presence of radiation (both photon and neutrino). 

In particular, in SN/GRB central engines, and also in putative reconnection layers in magnetar flares (see section~\ref{subsec-reconnection}), we are dealing with extreme environments that are very dense and relativistically hot ($k_B T > m_{\rm e} c^2$).  This means that, even though the plasma inertia may be dominated by the baryons (in the SN/GRB case), the particle number density, the pressure, and the entropy are all dominated by a hot closely-coupled plasma of pairs and photons in local thermodynamic equilibrium (LTE).  The electron, positron, and photon numbers are not conserved but instead are constantly maintained by pair and photon creation and annihilation processes near the levels dictated by the LTE condition, which can be expressed as an analog of the Saha equilibrium.  The densities and pressures then essentially become unique functions of the temperature and the lepton fraction (which in the SN/GRB environments is affected by the deleptonization processes associated with neutrino losses).  As a result, all the thermodynamic properties of such a plasma, including the effective equation of state, are significantly altered relative to the more familiar standard plasmas where no particles are created or destroyed.  

Furthermore, radiation controls most of the transport processes in these environments --- thermal conduction, resistivity, viscosity.  Importantly, the radiative processes are strongly affected by ultra-strong magnetic fields as we discussed in section~\ref{subsec-light-propagation}. 
At relatively modest Thomson optical depths (perhaps across a current sheet in a magnetar magnetosphere) radaitive cooling by photon diffusion may be significant and may control the energy balance. 
Also, radiative Silk damping due to photon thermal conduction may affect sound waves, etc.. 
However, in the case of SN/GRB central engines, the  environments of interest are so dense and optically thick that the photon mean free path is negligibly short. Then, photon-mediated transport processes (thermal conduction, viscosity,  etc.) can be completely ignored on the length- and time-scales dictated by the global MHD dynamics.  The same is also true for transport due to electrons, positrons, ions, and, under some circumstances, neutrons.

However, this does not necessarily mean that the plasma in SN and GRB central engines can be described by simple ideal MHD on the time scales of interest.  The reason for this is that the role of photon radiative energy transport  is taken up by neutrinos.  Neutrino cooling and heating can play a dominant role in governing the thermodynamics of these systems. A truthful description of neutrino transport is complicated by the large number of different neutrino processes that may need to be taken into account, depending on the conditions. These processes include neutrino-neutrino scattering; neutrino-antineutrino annihilation; neutrino-electron (and positron) scattering; neutral-current neutrino scattering off of free nucleons, $\alpha$ particles, and nuclei; charged-current neutrino and anti-neutrino absorption and emission by free nucleons and nuclei, and so on [see, e.g., \cite{Bruenn-1985, Reddy_etal-1998, Burrows_etal-2006b} for through reviews.]. 
Comprehensive modeling of neutrino transport processes in SN/GRB central engines is further complicated by the substantial range of neutrino optical depths that need to be covered.  Whereas inside NSs and in the densest parts of collapsar accretion disks neutrinos are optically thick and one can employ the radiation diffusion approximation, the surrounding areas (e.g., the post-bounce shock region in SN or at the base of the jet in GRBs) are optically thin to neutrinos which hence leak through them easily.  These lower-density regions are actually very important:  it is here that one would like to see the neutrinos that come out from the hotter denser central regions to deposit enough of their energy to revive a SN shock or to create a high-entropy photo-leptonic fireball to produce a GRB explosion.  

Furthermore, since the cross-sections of various reactions between neutrinos and normal matter are often rather strong functions of the neutrino energy, one has to treat neutrinos of different energies differently.  In particular, the boundary between optically thick and optically thin regions -- the neutrino-sphere --- may strongly depend on the neutrino energy. 
Perhaps the only feature of neutrino  transport that may make its detailed modeling simpler than, say, photon transport in a stellar atmosphere is the fact that neutrino cross-sections are smooth functions of energy, with no small-scale energy structure (i.e., the absence of spectral lines), for which exact or approximate analytical expressions are available~\cite{Bruenn-1985, Reddy_etal-1998, Burrows_etal-2006b}. 

Finally, nuclear composition of extreme astrophysical environments of core-collapse SN/GRB central engines is not fixed and can be far from that dictated by nuclear statistical equilibrium. It often has to be evolved self-consistently via a nontrivial network of nuclear reactions, which also  affects the thermodynamics of the plasma. 

These remarks point to the desirability of developing more realistic theoretical and numerical models of basic astrophysical MHD processes that pay a more careful attention to the questions of thermodynamics. One needs to go beyond the basic purely dynamical models that treat heating and cooling in a greatly simplified way, e.g.,  with a rudimentary equation of state.


\subsection{Plasma-surface interaction in magnetar magnetosphere}
\label{subsec-surface}

As discussed in section~\ref{subsec-magnetars}, the magnetic field outside an active magnetar is dynamic; both persistent X-ray emission and SGR gamma-ray flares are believed to be powered by the dissipative decay of the electrical currents induced in the force-free magnetosphere by the footpoint displacements on the NS surface. 
In particular, the question of how the electron-positron plasma carrying these currents through the magnetosphere interacts with the surface becomes an important and nontrivial issue (e.g., \cite{Beloborodov_Thompson-2007}).  As the intense high-energy relativistic particle beam impinges on the neutron star's crust, it is stopped over a relatively short distance. The resulting dissipation of the beam's energy heats up the star's surface locally, perhaps melting the crust,  and powers an intense hard X-ray emission in AXPs.  Morphologically, this process is similar to two-ribbon solar flares, in which the hard X-ray and white-light emission from the solar surface is powered by high-energy electrons accelerated by reconnection in the corona high above the surface of the Sun and streaming down along the reconnected magnetic field lines.  The underlying plasma physics of the beam-surface interaction, however, may be quite different in the two cases.  In the case of magnetars, this interaction may involve collective collisionless plasma processes, such as the excitation and subsequent dissipation of small-scale microscopic Langmuir and ion-acoustic turbulence by kinetic plasma instabilities~\cite{Beloborodov_Thompson-2007}. 
As discussed in section~\ref{subsec-B_*-physics},  under the conditions of interest, these processes take place on Compton (QED) scales and may be affected by  the presence of an ultra-strong magnetic field.  This means that their proper treatment and analysis requires developing a good understanding of relativistic quantum plasma physics, the foundations of which are outlined in section~\ref{sec-Quantum-Plasma-Physics}. 



\subsection{Plasma physics of reconnection of ultra-strong magnetic fields in SGR flares and GRB central engines} 
\label{subsec-reconnection}

Magnetic reconnection is a fundamental plasma process involving a rapid change of magnetic topology and often leading to a violent release of magnetic energy. 
How reconnection occurs, how fast it is, what its onset conditions are, and what are its outcomes, are all fundamental plasma-physical questions that have direct astrophysical implications. 
Magnetic reconnection of magnetar-strength fields ($B > B_*\simeq 4\times 10^{13}$~G) is one of the most important extreme plasma astrophysics problems with applications to magnetar magnetospheres and GRB central engines and jets (see sections~\ref{subsec-magnetars}-\ref{subsec-SN_GRB} and~\ref{subsec-MHD}).  
It is the most dramatic example of high-energy-density (HED) astrophysical reconnection.  
The physical conditions in the hot, dense lepto-photonic plasma in an ultra-magnetizied reconnection layer  are very different from those in the conventional, low-density heliospheric plasmas.  
A rich variety of exotic physical effects (e.g., special relativity, radiative effects, pair creation,  and QED effects due to the ultra-strong field) come into play in a most powerful way in this relativistic quantum plasma, making this problem intellectually exciting.  
Understanding the basic physics relevant to this problem represents a new open frontier in magnetic reconnection research. 
Perhaps the main goals of ultra-strong magnetic field reconnection research are to establish how rapidly and in what  form the dissipated energy emerges from the reconnection region. This includes calculating potentially observable radiative characteristics of the reconnection layer, such as its photospheric temperature and flare duration.  Comparing the results of such calculations with observations would enable one to assess whether, for example, reconnection is a plausible mechanism for SGR giant flares and for GRBs. 

Developing a detailed quantitative picture of this process, with all the bells and whistles of QED in an ultra-strong field, has not yet been accomplished.  Generally speaking, this is a multi-scale phenomenon, with scales ranging from the Compton scale~$l_C=\hbar/m_e c$, to the scales relevant to the radiative transfer in, and the growth of, the optically thick pair coat described below (from the photon mean free path $l_{\rm mfp} \sim \alpha^{-3}\, l_C$ to the photosphere thickness~$\delta_{\rm ph}$), to the astrophysical scales of relevance (neutron star radius $R_{\rm NS} \sim 10^6$~cm).

As a first step towards understanding reconnection of ultra-strong fields, Uzdensky \cite{Uzdensky-2011}  made simple preliminary estimates that enable one to identify the key physical ingredients of the process.  This work should serve as the basis for any rigorous theoretical model of this process. 
The key findings can be summarized as follows.
When the energy of an ultra-strong magnetic field is dissipated, the resulting internal plasma energy density is so high that the plasma inside the reconnection region becomes relativistically hot and one expects copious production of photons and electron-positron pairs in local thermodynamic equilibrium (LTE) with each other (see section~\ref{subsec-B_*-physics}).
The temperature~$T_0$ and the equilibrium pair density $n_{\rm pairs}$  at the center of the reconnection layer can be estimated in terms of the reconnecting magnetic field $B_0$ from the cross-layer pressure balance:  $P_{\rm rad, 0} + P_{\rm pairs, 0} = B_0^2/8\pi$, where $P_{\rm rad, 0}=  a T_0^4 /3 = (\pi^2/45)\, [m_{\rm e} c^2/(l_{\rm C}^3)] \, \theta_{\rm e}^4 $ 
and $P_{\rm pairs, 0} \simeq (7/4) \, P_{\rm rad}$  are the layer's radiation pressure and the relativistic pair plasma pressure, respectively.  One then obtains 
$\theta_{\rm e} \equiv k_{\rm B} T_0/m_{\rm e} c^2 \simeq (45/22\,\pi^3\alpha)^{1/4}\, b^{1/2}  \simeq 1.73\, b^{1/2} $, 
where, once again, $b \equiv B_0/B_* $ and $\alpha$ is the fine structure constant.
Correspondingly, the pair density inside the layer, determined by the LTE equilibrium condition in terms of $T_0$, becomes (in the ultra-relativistic limit $\theta \gg 1$ corresponding to $b\gg 1$)
\beq
n_{\rm pairs} \simeq 0.183\, (k_{\rm B} T/{\hbar c})^3 \simeq
0.183\ l_{\rm C}^{-3}\, \theta_{\rm e}^3 \simeq 3.2\times 10^{30}\ {\rm cm}^{-3}\, \theta_{\rm e}^3 \, .
\label{eq-n_pairs}
\eeq
Thus, the characteristic density scale in an extreme reconnection layer is one particle per~$l_{\rm C}^3$.  This density is much higher than, and hence essentially independent from, the ambient plasma density in a magnetar magnetosphere. This justifies our assumption that the baryonic contribution to the pressure inside the layer should be negligible in the context of SGR flares.

The above very high pair plasma density implies that the global reconnection layer of length~$L\sim R_{\rm NS} \sim 10^6\, {\rm cm}$ is highly collisional  in the traditional sense that the corresponding Sweet--Parker layer thickness $\delta_{\rm SP}(L)= \sqrt{L\eta/V_{\rm A}}$, where $\eta$ is the magnetic diffusivity and $V_{\rm A}$ is the Alfv\'en velocity, is much larger than the characteristic collisionless length scales, such as the electron skin depth~$d_e$, or the electron Larmor radius~$\rho_e$~\cite{Uzdensky_MacFadyen-2006, Uzdensky-2011, McKinney_Uzdensky-2012}. 
This means that the classical collisional and Compton-drag resistivities should dominate over other non-ideal terms in the generalized Ohm law and hence that collisional radiative resistive MHD description should be valid on this scale. 

However, since the corresponding global Lundquist number~$S=L V_{\rm A}/\eta$ for this environment is huge, $S \sim 10^{18}$ \cite{Uzdensky-2011}, much higher than the critical number $S_{\rm c} \sim 10^4$ for the onset of the secondary tearing (aka plasmoid) instability~\cite{Loureiro_etal-2007, Loureiro_etal-2013}, the global Sweet-Parker layer is violently tearing-unstable~\cite{Samtaney_etal-2009}.  Then, reconnection proceeds in the highly dynamic plasmoid-dominated regime, via a hierarchical chain of secondary plasmoids and current sheets \cite{Shibata_Tanuma-2001, Daughton_etal-2009, Samtaney_etal-2009, Bhattacharjee_etal-2009, Huang_Bhattacharjee-2010, Uzdensky_etal-2010, Loureiro_etal-2012}. This  provides a path towards forming very small dissipative structures --- the elementary current sheets at the bottom of the hierarchy --- that develop on microscopic plasma-physical scales, i.e., the Larmor radius or the collisionless skin depth (which, as we saw in section~\ref{subsec-B_*-physics}, are on the Compton length-scale) or the critical Sweet--Parker scale $\delta_{\rm c} \simeq 100 \, \eta/V_{\rm A}$ \cite{Uzdensky_etal-2010}.  This suggests that even in the very dense and highly collisional extreme astrophysical environments considered in this review, detailed plasma physics beyond the standard resistive MHD may still be important in controlling energy dissipation processes.  
If the classical conventional reconnection theory can serve us as a guide, the effective reconnection rate should be bracketed between the purely resistive-MHD plasmoid-regime rate of~$0.01\, B_0 V_{\rm A}$ and the often-quoted canonical collisionless reconnection rate of~$0.1\, B_0 V_{\rm A}$ \cite{Uzdensky-2011}.  Which one of these limits applies should  depend on the ratio $\delta_{\rm c}/d_{\rm e} \sim 100\, \eta V_{\rm A}/d_{\rm e}$, and preliminary estimates suggest that, for the physical conditions of interest, these two scales, $\delta_{\rm c}$ and~$d_{\rm e}$, are roughly comparable.  
Whether this general picture holds, and how it is modified by the QED effects, intense radiation, and ultra-strong magnetic fields, is not at all clear.  This calls for the development of a more precise physical description of microscopic current layers on the Compton scale, e.g., calculating plasma resistivity and other non-ideal-MHD effects,  based on relativistic quantum plasma theory.  

The above reconnection rate range ($0.01-0.1\, B_0 V_{\rm A}$) has strong implications for magnetar flares~\cite{Uzdensky-2011}.  It translates into a very short reconnection time of about~$1\, {\rm msec}$, much shorter than the typical observed duration of the main spike (~ 0.25 sec). This implies that the reconnection process itself is {\it not} the main energy-release bottleneck and that it does not govern the flare duration.  Instead, the duration may be governed by the escape of radiation from the hot fireball ejected out of the reconnection region, and/or by the slow driving by the neutron star crustal motions~\cite{Thompson_Duncan-1995, Thompson_Duncan-2001, Lyutikov-2006a}.

The second important role of the pairs produced in the reconnection process is their effect on the propagation of photons out of the layer, which affects both the radiative cooling of the layer and its observational appearance.  The former is important because the equilibrium pair density, resistivity, and radiation pressure are all strong functions of temperature, and hence the issues of thermodynamics and thermal transport are critical. 

In the case of reconnection of magnetar-strength fields, the compactness of the reconnection region is very large. Therefore, the high pair density is not  limited to the thin current layer proper in which ohmic dissipation of the magnetic energy takes place.  Instead, copious pair production by high-energy $\gamma$-ray photons emitted by the layer quickly dresses it in a continuously growing coat of electron-positron pairs~\cite{Uzdensky-2011}.  As it grows, this pair coat may become much thicker than the current layer proper, and thus may appear to an outside observer as a somewhat flattened fireball. 
The coat becomes optically thick to Compton scattering and to 1-photon pair creation and photon-splitting, even in the direction across the layer.  This ensures good local thermal coupling between the photons and the plasma ---  in stark contrast with conventional low-energy-density reconnection found in, e.g., solar flares or the Earth magnetosphere. 

The reconnection problem then becomes, fundamentally, a time-dependent {\it radiative transfer problem}; this aspect is essential to understanding the energetics of reconnection in this environment.
On length scales much larger than the photon mean free path~$\lambda_{\rm ph}$, but smaller than the thickness of the growing pair coat, the problem can be treated in the radiation diffusion approximation. 
The radiative transfer equations are supplemented by the expressions for the local pair density and the photon mean-free-path~$\lambda_{\rm ph}$.  In most of the pair coat, deep below the photosphere, the plasma-radiation thermal coupling is good and the LTE density is determined by an analog of the Saha formula; it is a strong function of the local temperature.  The photon mean free path, then, also becomes a strong function of the temperature but, in addition, is greatly affected by the ultra-strong magnetic field through the QED effects described in section~\ref{subsec-light-propagation}. 
Near the pair coat's photosphere, the situation becomes even more complicated: the  LTE and radiative diffusion approximations both break down and so one needs to build a numerical kinetic model of radiative and pair-production processes taking place here.

Because of the large optical depth~$\tau$, photons diffuse gradually across the dressed layer of thickness~$H$. However, in the case of a moderately subcritical reconnecting field ($B_0 \ll B_*$) and hence a sub-relativistic layer temperature and moderate coat optical depth, the photon diffusion time $t_{\rm diff} \sim \tau H/c$ may be shorter than the global advection time $t_{\rm adv} \sim L/c$ along layer of length~$L$.  Then, photons can escape the layer before being advected out by the plasma flow if $\tau < L/H$.  In this case, a significant fraction of the dissipated magnetic energy promptly leaves the layer by rapid radiative diffusion across it, instead of being advected out along the layer as in traditional reconnection. 
In this {\it strong cooling} regime~\cite{Uzdensky_McKinney-2011}, radiative cooling becomes the main mechanism of removing the dissipated energy; it governs the thermodynamics of the layer and drastically affects its appearance to an external observer.  In particular, in the case of steady state reconnection without guide field, one can show that the optical depth across the dressed layer~$\tau$, and the reconnection rate~$cE= v_{\rm rec} B_0$ are related simply by
$\tau \simeq T_0^4/T_{\rm ph}^4 \sim B_0/E \sim c/v_{\rm rec}$~\cite{Uzdensky-2011}. 
Thus, the thicker the layer, the slower the reconnection process. 

It turns out, however,  that the strong cooling assumption (i.e., a stationary balance between reconnection ohmic heating and radiative cooling)  is not valid in the case of super-critical, magnetar-strength magnetic fields~\cite{Uzdensky-2011}. The reason for this is that the central temperature in the layer is expected to be $\sim 1.7\, b^{1/2}\, m_{\rm e} c^2$ (about 3~MeV for $B\sim 10 B_*$), whereas the photosphere of the pair coat forms at $k_B T_{\rm ph} \simeq 20-30$ keV (because $\lambda_{\rm ph,mfp}$ due to Thomson scattering in an equilibrium pair plasma becomes comparable to the global system size $L\sim R_{\rm NS} \sim 10^6$~cm at this temperature).  In a steady state, this would imply a huge optical depth of order $\tau\sim T_0^4/ T_{\rm ph}^4 > 10^5\, b^2$, and then the radiative diffusion time would be much longer than both the observed flare time and the global advection time along the layer, $t_{\rm adv} \sim L/c$. 
This means that the reconnection-powered growth of the pair coat is not in the stationary strong-cooling regime and has to be considered as a time-dependent process.  In the context of SGR flares, most of the magnetic energy dissipated in a reconnection event directly powers the continuous production of pairs and photons that make up the growing pair coat.  The reconnection process can thus be viewed as an efficient conversion of magnetic energy into the energy of a relativistically hot lepto-photonic fireball surrounding the reconnection layer. 
On time-scales of order the global Alfv\'en (or light-) advection time $t_{\rm adv} \sim L/c$, this newly created pair plasma gets torn apart into two halves by the large-scale reconnection outflow,  which directly leads to the formation of two optically thick, dense and hot plasma blobs, as discussed in section~\ref{subsubsec-SGR-flares}.  One of them can be associated with the ejected  fireball that powers the main SGR flare spike emission  and the other with the fireball trapped by the reconnected post-flare magnetic loops close to the star and responsible for the long-lasting tail emission.  


Some of the above considerations for reconnection in extreme ultra-magnetized environments  also have implications for the workings of central engines of SNe and GRB jets. In particular, the inferred reconnection rate may be important for the survival of the inner parts of GRB jets \cite{Uzdensky_MacFadyen-2006, McKinney_Uzdensky-2012}. 
Deep inside the star, where the plasma in a reconnection layer is highly collisional, reconnection might not be fast enough to  destroy the magnetic jet, which then may be able to survive the propagation to larger distances~\cite{Uzdensky_MacFadyen-2006}.  Eventually, well outside the star, the magnetic field in the jet drops, reconnection layers cool, pairs annihilate, and the plasma becomes optically thin and collisionless. 
Then, an ongoing reconnection process can switch to the fast collisionless reconnection regime, leading to a renewed magnetic energy release~\cite{McKinney_Uzdensky-2012}. 
For  typical collapsar parameters, McKinney and Uzdensky \cite{McKinney_Uzdensky-2012} found that this collisionality-induced reconnection switch is expected to trigger the transition to the fast reconnection 
at a relatively large distance ($r\sim 10^{13}$--$10^{14}{\rm cm}$), giving the flow enough range to accelerate to a high Lorentz factor ($\gamma\sim 100$--$1000$).
It was concluded that this reconnection switch mechanism allows for efficient conversion of electromagnetic energy into prompt emission  with radiation signatures consistent with observations. 



\section{Summary}
\label{sec-summary}

In this review we focused on a rather exotic branch of plasma astrophysics --- namely, the plasma physics of what we called extreme astrophysical environments --- ultra-magnetized systems with magnetic fields in excess of the critical quantum field $B_* \simeq 4.4 \times 10^{13}\, {\rm G}$. 

We started with a brief introduction outlining our general view of plasma astrophysics as a study of  physical processes  responsible for energy exchange between various constituents of complex cosmic plasmas, such as thermal gas, nonthermal particles,  large-scale magnetic field, small-scale turbulent motions, radiation, etc.  (section~\ref{sec-intro}).  After this general discussion we concentrated on the extreme environments of interest to this review.  We presented a broad phenomenological overview of astrophysical systems in which such conditions are encountered in section~\ref{sec-extreme-astro-environments}.  Throughout this section we tried to identify important plasma-physical problems that need to be resolved in order for us to gain a good understanding of how these systems work. 
The systems in question are relatively rare and unique, but they are behind some of the most spectacular high-energy phenomena observed in the modern Universe.  We first talked about magnetic fields in neutron stars, with a focus on the problem of neutron star magnetogenesis and on the subsequent evolution of magnetic field in neutron stars (section~\ref{subsec-NS-interior}).  We then considered what can perhaps be viewed as the most clear, unambiguous astrophysical example of systems with a supercritical magnetic field --- active magnetospheres of magnetars, neutron stars with large-scale magnetic fields of order $10^{15}\, {\rm G}$ (section~\ref{subsec-magnetars}).  Observationally, magnetars come in two classes, AXPs and SGRs, and we discussed both the powerful persistent X-ray emission observed in both classes (section~\ref{subsubsec-magnetars-Xray})  and also the remarkable giant $\gamma$-ray flares that represent the hallmark of SGR activity (section~\ref{subsubsec-SGR-flares}).  Both of these phenomena are believed to be powered by the magnetic activity in neutron star magnetospheres, which may have many parallels to the activity in the solar corona.  
After discussing magnetar magnetospheres, we devoted section~\ref{subsec-SN_GRB} to a basic introduction to systems that have even more extreme conditions (e.g., energy densities) --- the central engines of core-collapse SNe and long GRBs.  Since the main emphasis of this article is on plasma-physical aspects of the relevant astrophysical systems, we focused our discussion on the role of MHD (e.g., magnetorotational) processes in powering these explosions and driving GRB jets. 
We note that, while we did not include Type Ia supernovae and short GRBs in our discussion,  we believe much of the physics discussed in this review should also be relevant to these systems.

The numerous plasma-physical aspects of the exotic systems described in section~\ref{sec-extreme-astro-environments} underlie the need to develop a coherent theoretical framework that would be appropriate for correctly describing various phenomena in ultra-magnetized, often relativistically hot plasmas.  Many of the relevant plasma processes, e.g., wave propagation, turbulence, reconnection,  are superficially similar to well-studied processes encountered in traditional, low-energy-density,  space- and astrophysical contexts. 
It is important to recognize, however,  that the basic physical conditions in the above extreme astrophysical systems are very different from those in low-energy-density plasmas, and so many familiar plasma processes are strongly modified.  In particular, as we explained in section~\ref{subsec-B_*-physics}, the presence of an ultra-strong magnetic field inevitably calls for a relativistic quantum description of these plasmas.  While QED-based relativistic quantum plasma physics in ultra-strong magnetic fields is still much less developed than classical plasma physics, a lot of important fundamental work has actually been done in this area over the past few decades.  We reviewed this work and outlined the key conceptual steps toward constructing relativistic quantum plasma physics in section~\ref{sec-Quantum-Plasma-Physics}. 
Basic QED properties of supercritical magnetic fields, including single-particle motion, basic properties of a pair plasma with energy density of $B_*^2/8\pi$, the QED phenomenon of vacuum polarization and its effect on the propagation of electromagnetic waves in vacuum, and the equilibrium statistical mechanics of a fermionic gas immersed in such a field, were reviewed in section~\ref{subsec-B_*-physics}. 
Section~\ref{subsec-nonrel-QP} was devoted to an exposition of non-relativistic quantum plasma physics.  We started with the Wigner-Moyal formalism for quantum kinetic theory (section~\ref{subsubsec-Wigner-Moyal}), then described the application of this formalism to linear waves and instabilties (section~\ref{subsubsec-quantum-waves}),  discussed the efforts to develop quantum fluid theories, such as QMHD (section~\ref{subsubsec-nonrel-quantum-fluid}), and to incorporate spin effects in the quantum fluid theory (Pauli plasma, see section~\ref{subsubsec-nonrel-spin-plasma}), and, finally, gave a brief account of the recent progress in studying nonlinear plasma phenomena in non-relativistic quantum plasmas (section~\ref{subsubsec-nonrel-nonlinear}). 
We then moved on to relativistic quantum plasma physics (section~\ref{subsec-rel-QP}), which is developed even less than its nonrelativistic counterpart. This theory is properly built starting with QED as its foundation.  Most of the work done in this area so far falls into two categories: physics of plasmas consisting of spinless charged particles (Klein-Gordon plasma, section~\ref{subsubsec-KG}) and of those consisting of spin-1/2 fermions (Dirac plasma, section~\ref{subsubsec-Dirac}).  We also discussed relativistic quantum hydrodynamics in section~\ref{subsubsec-rel-hydro}. 

In section~\ref{sec-astro-applications}, we returned to the astrophysical applications of extreme plasma physics. Here we considered several outstanding plasma-physical problems that naturally arise from the discussion of ultra-magnetized astrophysical objects presented in section~\ref{sec-extreme-astro-environments}.  We  discussed: (1) propagation of high-energy photons in a magnetar magnetosphere (section~\ref{subsec-light-propagation});  
(2) large-scale MHD processes governing magnetar activity, magnetic field generation in newly-born proto-neutron stars and in collapsar accretion disks, formation and propagation of Poynting-flux-dominated magnetic tower jets in core-collapse GRB central engines, and MHD instabilties and current-sheet formation expected during the propagation of these jets to large distances (section~\ref{subsec-MHD}); 
(3) various thermodynamic aspects of plasmas in extreme high-energy-density, radiation-dominated environments (section~\ref{subsec-thermo}); 
(4) interaction between intense currents flowing through a twisted magnetar magnetosphere and the magnetar surface, presumably involving the excitation and dissipation of plasma micro-turbulence and believed to be responsible for the observed persistent X-ray emission in AXPs and SGRs (section~\ref{subsec-surface}); 
and, finally, 
(5) magnetic reconnection involving ultra-strong magnetic fields in the context of magnetar flares and GRB central engines (section~\ref{subsec-reconnection}). 

In conclusion, we hope that this review will motivate further theoretical work to develop relativistic quantum plasma physics from first principles of QED, as well its applications to real astrophysical problems with the goal of gaining a more rigorous, physics-based understanding of ultra-magnetized neutron stars and related systems.  We finally note that while most of the work on quantum and relativistic quantum plasma physics has so far been purely analytical, at some point, once the theoretical foundation has been firmly established, the research focus should shift towards developing numerical methods suitable for attacking more complex, nonlinear problems that are more directly relevant to real systems.  In classical plasma physics such a shift already occurred a couple decades ago, and we believe that one should expect a similar shift in (relativistic) quantum plasma physics in the near future, once the discipline matures.


\ack

We are grateful to A. Beloborodov, M. Lyutikov, J. McKinney, and M. Medvedev for useful discussions. 
This work was supported by 
the US National Science Foundation grant  PHY-0903851 
and  Department of Energy grants DE-SC0008409 and DE-SC0008655

\section*{References}
\bibliographystyle{unsrt}

\bibliography{rop}

\end{document}